\documentclass[gmdd,manuscript]{copernicus}

\begin{document}

\title{The Modular Arbitrary-Order Ocean-Atmosphere Model: \textsc{maooam}~v1.0}

\Author{Lesley}{De~Cruz}
\Author{Jonathan}{Demaeyer}
\Author{St\'{e}phane}{Vannitsem}

\affil{Royal Meteorological Institute of Belgium, Avenue Circulaire 3, 1180 Brussels, Belgium}

\correspondence{Lesley De~Cruz (lesley.decruz@meteo.be)}

\runningtitle{\textsc{maooam}}
\runningauthor{L.~De~Cruz et al.}

\received{}
\pubdiscuss{} 
\revised{}
\accepted{}
\published{}

\nolinenumbers 

\firstpage{1}

\maketitle

\begin{abstract}
This paper describes a reduced-order quasi-geostrophic coupled
ocean--atmosphere model that allows for an arbitrary number of atmospheric
and oceanic modes to be retained in the spectral decomposition. The
modularity of this new model allows one to easily modify the model physics.
Using this new model, coined the ``Modular Arbitrary-Order Ocean-Atmosphere
Model'' (\textsc{maooam}), we analyse the dependence of the model dynamics on
the truncation level of the spectral expansion, and unveil spurious behaviour
that may exist at low resolution by a comparison with the higher-resolution
configurations. In particular, we assess the robustness of the coupled
low-frequency variability when the number of modes is increased. An
``optimal'' configuration is proposed for which the ocean resolution is
sufficiently high, while the total number of modes is small enough to allow
for a tractable and extensive analysis of the dynamics.
\end{abstract}

\introduction

The atmosphere at mid-latitudes displays a variability on a wide range of
space scales and timescales, and in particular a low-frequency variability at
interannual and decadal timescales as suggested by the analyses of different
time series developed in the past years
\citep{Trenberth1990,TH1994,Hurrell1995,Mantua1997,LW2003,LS2013}. In
contrast to the phenomenon of El Ni\~no--Southern Oscillation (ENSO), of
which the driving mechanisms are intensively studied and quite well
understood \citep[e.g.][]{Philander1990, GZ2013}, the origin of mid-latitude
low-frequency variability (LFV) remains highly debated, mainly due to the
poor ability of state-of-the-art coupled ocean--atmosphere models to simulate
it correctly \citep[e.g.][]{NLA2011,SSEK2014}. The most plausible candidates
of this LFV are either the coupling with the local ocean \citep{KDBMG2007},
or teleconnections with the tropical Pacific ocean--atmosphere variability
\citep{MFC2008}, or both.

Recently the impact of the coupling between the ocean and the atmosphere at
mid-latitudes on the atmospheric predictability \citep{ND1993,R1995,PK2004}
and the development of the LFV \citep{V2003} has been explored in a series of
low-order coupled ocean--atmosphere systems. However, the limited flexibility
of the possible geometries of these previous models led the present authors
to develop a series of new model versions. The first of these, OA-QG-WS~v1
\citep{vannitsem2014}, for Ocean-Atmosphere--Quasi-Geostrophic--Wind Stress,
features only mechanical coupling between the ocean and the atmosphere, and
uses 12 atmospheric variables following \citet{CS1980} and four oceanic modes
following \citet{P2011}. In a successor of this model, OA-QG-WS~v2, the set
of atmospheric variables is extended from 12 to 20 as in \citet{RP1982}. This
increase in resolution in the atmosphere was shown to be key to the
development of a realistic double gyre in the ocean \citep{VD2014}. A third
version of this model, hereafter referred to as \textsc{vddg} in reference to
the authors of the model, includes passively advected temperature in the
ocean and an energy balance scheme, combined with an extended set of modes
for the ocean \citep{VDDG2015}.

In the \textsc{vddg} model, an LFV associated with the coupling between the
ocean and the atmosphere is successfully identified, allowing for
extended-range coupled ocean--atmosphere predictions. Moreover, the
development of this coupled ocean--atmosphere mode is robust when stochastic
forcings are added \citep{DV2016}, or when a seasonal radiative forcing is
incorporated into the low-order model \citep{V2015}. Remarkably the presence
of the seasonal radiative input favours the development of the coupled mode
due to the amplification of the impact of the wind stress forcing in summer,
associated with a drastic reduction of the mixed layer thickness at that
period of the year. While these are encouraging results, which suggest the
generic character of the coupled ocean--atmosphere mode, they need to be
confirmed through the analysis of more sophisticated models, and in
particular in higher-resolution coupled systems.

In this article, we present a model that generalizes the \textsc{vddg} model
by allowing for an arbitrary number of modes, or basis functions in which the
dynamical fields are expanded. The modes can be selected independently for
the ocean and the atmosphere, and for the zonal and meridional directions.
The modular approach allows one to straightforwardly modify the model
physics, such as changing the drag coefficient, introducing new dissipative
schemes or adding a seasonal insolation. This model was coined
\textsc{maooam}: the Modular Arbitrary-Order Ocean-Atmosphere Model. The
model equations and its technical implementation are detailed in Sect.~2. In
Sect.~3, \textsc{maooam} is used to investigate the dependence of the model
dynamics, i.e. its climatology and the qualitative structure of its
attractor, on the number of modes included. Furthermore, the development of
the LFV as a function of the spectral truncation is discussed. Key results
are summarized in Sect.~4.

\section{Model formulation}\label{sec:model}

The model is composed of a two-layer quasi-geostrophic (QG) atmosphere,
coupled both thermally and mechanically to a QG shallow-water ocean layer, in
the $\beta$-plane approximation. The atmospheric component is an extension of
the QG model, first developed by \citet{CS1980} and further refined by
\citet{RP1982}. The equations of motion for the atmospheric streamfunction
fields $\psi^1_\text{a}$ at 250\,hPa and $\psi^3_\text{a}$ at 750\,hPa, and
the vertical velocity $\omega = \text{d}p/\text{d}t$, read
\begin{align}
\frac{\partial}{\partial t} \left( \nabla^2 \psi^1_\text{a} \right) +
J(\psi^1_\text{a}, \nabla^2 \psi^1_\text{a}) + \beta \frac{\partial
  \psi^1_\text{a}}{\partial x} \label{eqn:atmos1} &=  -k'_d \nabla^2 (\psi^1_\text{a}-\psi^3_\text{a}) + \frac{f_0}{\Delta p} \omega,\\
\frac{\partial}{\partial t} \left( \nabla^2 \psi^3_\text{a} \right) +
J(\psi^3_\text{a}, \nabla^2 \psi^3_\text{a}) + \beta \frac{\partial
  \psi^3_\text{a}}{\partial x} \label{eqn:atmos3} &= +k'_d \nabla^2 (\psi^1_\text{a}-\psi^3_\text{a}) - \frac{f_0}{\Delta
p} \omega - k_d \nabla^2 (\psi^3_\text{a}-\psi_\text{o}).
\end{align}

The Coriolis parameter $f$ is linearized around a value $f_0$ estimated at
latitude $\phi_0 = 45${\degree}\,N, $f = f_0 + \beta y$, with
$\beta=\text{d}f/\text{d}y$. The parameters $k'_d$ and $k_d$ quantify the
friction between the two atmospheric layers and between the ocean and the
atmosphere, respectively, and $\Delta p = 500$\,hPa is the pressure
difference between the atmospheric layers.

The equation of motion for the streamfunction $\psi_\text{o}$ of the ocean
layer reads \citep[cf.][]{P2011}
\begin{align}
\frac{\partial}{\partial t} \left( \nabla^2 \psi_\text{o} -
\frac{\psi_\text{o}}{L_\text{R}^2} \right) + J(\psi_\text{o}, \nabla^2
\psi_\text{o}) + \beta \frac{\partial \psi_\text{o}}{\partial x}
\label{eqn:ocean}
& = -r \nabla^2 \psi_\text{o} +\frac{C}{\rho h} \nabla^2
(\psi^3_\text{a}-\psi_\text{o}).
\end{align}
$L_\text{R}$ is the reduced Rossby deformation radius, $\rho$ the density,
$h$ the depth, and $r$ the friction at the bottom of the active ocean layer.
The rightmost term represents the impact of the wind stress, and is modulated
by the drag coefficient of the mechanical ocean--atmosphere coupling, $d =
C/(\rho h)$.

The time evolution of the atmosphere and ocean temperatures $T_\text{a}$ and
$T_\text{o}$ obeys the following equations:
\begin{align}
\gamma_\text{a} \left( \frac{\partial T_\text{a}}{\partial t} + J(\psi_\text{a}, T_\text{a}) -\sigma \omega \frac{p}{R}\right) \label{eqn:heat_atmos}
& = -\lambda (T_\text{a}-T_\text{o}) + \epsilon_\text{a} \sigma_\text{B} T_\text{o}^4 - 2 \epsilon_\text{a} \sigma_\text{B} T_\text{a}^4 + R_\text{a}, \\
\gamma_\text{o} \left( \frac{\partial
T_\text{o}}{\partial t} + J(\psi_\text{o}, T_\text{o}) \right) \label{eqn:heat_ocean}
  & = -\lambda (T_\text{o}-T_\text{a}) -\sigma_\text{B} T_\text{o}^4 +
\epsilon_\text{a} \sigma_\text{B} T_\text{a}^4 + R_\text{o}.
\end{align}
Here, $\gamma_\text{a}$ and $\gamma_\text{o}$ are the heat capacities of the
atmosphere and the active ocean layer.
$\psi_\text{a}=(\psi^1_\text{a}+\psi^3_\text{a})/2$ is the atmospheric
barotropic streamfunction. $\lambda$ is the heat transfer coefficient at the
ocean--atmosphere interface, and $\sigma$ is the static stability of the
atmosphere, taken to be constant. The quartic terms represent the long-wave
radiation fluxes between the ocean, the atmosphere, and outer space, with
$\epsilon_\text{a}$ the emissivity of the grey-body atmosphere and
$\sigma_\text{B}$ the Stefan--Boltzmann constant. By decomposing the
temperatures as $T_\text{a} = T_\text{a}^0 + \delta T_\text{a}$ and
$T_\text{o} = T_\text{o}^0 + \delta T_\text{o}$, the quartic terms are
linearized around spatially uniform temperatures $T_\text{a}^0$ and
$T_\text{o}^0$, as detailed in Appendix~B of \citet{VDDG2015}. $R_\text{a}$
and $R_\text{o}$ are the short-wave radiation fluxes entering the atmosphere
and the ocean that are also decomposed as $R_\text{a}=R_\text{a}^0 + \delta
R_\text{a}$ and $R_\text{o} = R_\text{o}^0 + \delta R_\text{o}$.

The hydrostatic relation in pressure coordinates $(\partial \Phi/\partial p)
= -1/\rho_\text{a}$ where the geopotential height $\Phi^i =
f_0\;\psi_\text{a}^i$ and the ideal gas relation $p=\rho_\text{a} R
T_\text{a}$ allow one to write the spatially dependent atmospheric
temperature anomaly $\delta T_\text{a} = 2f_0\;\theta_\text{a} /R$, with
$\theta_\text{a} \equiv (\psi^1_\text{a}-\psi^3_\text{a})/2$ often referred
to as the baroclinic streamfunction. $R$ is the ideal gas constant. This can
be used to eliminate the vertical velocity $\omega$ from
Eqs.~(\ref{eqn:atmos1})--(\ref{eqn:atmos3}) and (\ref{eqn:heat_atmos}). This
reduces the independent dynamical fields to the streamfunction fields
$\psi_\text{a}$ and $\psi_\text{o}$, and the spatially dependent temperatures
$\delta T_\text{a}$ and $\delta T_\text{o}$.

The prognostic equations for these four fields are then non-dimensionalized
by dividing time by $f_0^{-1}$, distance by a characteristic length scale
$L$, pressure by the difference $\Delta p$, temperature by $f_0^2 L^2/R$, and
streamfunction by $L^2 f_0$. A more detailed discussion of the model
equations and their non-dimensionalization can be found in \citet{VD2014} and
\citet{VDDG2015}.

All the parameters of the model equations used in the present work are listed
in Table~\ref{tab:params}.

\subsection{Expansion of the dynamical fields}

In non-dimensionalized coordinates $x'=x/L$ and $y'=y/L$, the domain is
defined by $(0 \leq x' \leq \frac{2\pi}{n}, 0 \leq y' \leq \pi)$, with $n = 2
L_y/L_x$ the aspect ratio between its meridional and zonal extents (see
Table~\ref{tab:params} for the value used here). The atmospheric flow is
defined in a zonally periodic channel with no-flux boundary conditions in the
meridional direction ($\partial \cdot_\text{a}/\partial x' \equiv 0$ at $y'=
0,\pi$), whereas the oceanic flow is confined within an ocean basin by
imposing no-flux boundaries in both the meridional ($\partial
\cdot_\text{o}/\partial x' \equiv 0$ at $y'= 0,\pi$) and zonal ($\partial
\cdot_\text{o}/\partial y' \equiv 0$ at $x'= 0,2\pi/n$) directions. These
boundary conditions limit the functions used in the Fourier expansion of the
dynamical fields. With the proper normalization, the basis functions for the
atmosphere must be of the following form, following the nomenclature of
\citet{CT1987}:
\begin{align}
&F^A_{P} (x', y')   =  \sqrt{2}\, \cos(P y')\label{eqn:FA}\\
&F^K_{M,P} (x', y') =  2\cos(M nx')\, \sin(P y')\label{eqn:FK}\\
&F^L_{H,P} (x', y') = 2\sin(H nx')\, \sin(P y').\label{eqn:FL}
\end{align}

Analogously, the oceanic basis functions must be of the form
\begin{align}
\phi_{H_\text{o},P_\text{o}} (x', y') = & 2\sin(\frac{H_\text{o} n}{2} x')\,
\sin(P_\text{o} y'), \label{eqn:phi}
\end{align}
with integer values of $M$, $H$, $P$, $H_\text{o}$, and $P_\text{o}$.

For example, the spectral truncation used by \citet{CS1980} can be specified
as Eqs.~(\ref{eqn:FA})--(\ref{eqn:FL}) with $M=H=1$; $P \in \{1,2\}$.
\citet{RP1982} extend this set by two blocks of two functions each, and the
resulting set can be specified as $M,H \in \{1,2\}$; $P \in \{1,2\}$. The
\textsc{vddg} model has $M,H \in \{1,2\}$; $P \in \{1,2\}$ and $H_\text{o}
\in \{1,2\}$; $P_\text{o} \in \{1,2,3,4\}$. Note that, for consistency, the
ranges for $M$ and $H$ should be the same. The distinction between $M$ and
$H$ is, however, required to avoid ambiguities in the formulae of the inner
products, as specified in Appendix~\ref{app:formulae}.

For the given ranges of $1 \leq P \leq P^{\max}$, $1 \leq (M,H) \leq
H^{\max}$ and $1 \leq P_\text{o} \leq P_\text{o}^{\max}$, and $1 \leq
H_\text{o} \leq H_\text{o}^{\max}$, the number of basis functions can be
calculated as
\begin{align}
n_\text{a} =& P^{\max}\,(2\,H^{\max} + 1);\quad n_\text{o} =
P_\text{o}^{\max}\,H_\text{o}^{\max}.
\end{align}

Ordering the basis functions as in Eqs.~(\ref{eqn:FA})--(\ref{eqn:FL}), along
increasing values of $M= H_\text{(o)}$ and then $P_\text{(o)}$, allows one to
write the set as $\left\{ F_i(x',y'), \;\phi_j (x',y')\right\}$ $(1 \leq i
\leq n_\text{a}, 1 \leq j \leq n_\text{o} )$. The dynamical fields can then
be written as the following truncated series expansions:
\begin{align}
 \psi_\text{a} (x',y',t) &= \sum_{i=1}^{n_\text{a}} \; \psi_{\text{a},i}(t) F_i(x',y'), \label{eq:psidec}\\
 \delta T_\text{a}(x',y',t) &=\sum_{i=1}^{n_\text{a}} \delta T_{\text{a},i}(t) \; F_i(x',y'), \\
 &= 2 \frac{f_0}{R} \sum_{i=1}^{n_\text{a}} \theta_{\text{a},i}(t) \; F_i(x',y'), \nonumber\\
 \psi_\text{o}(x',y',t) &= \sum_{j=1}^{n_\text{o}} \psi_{\text{o},j}(t) \; (\phi_j(x',y') \; -\; \overline{\phi_j}), \label{eq:psiodec}\\
 \delta T_\text{o}(x',y',t) &= \sum_{j=1}^{n_\text{o}} \delta T_{\text{o},j}(t) \; \phi_j(x',y').\label{eq:Tdec}
\end{align}

Furthermore, the short-wave radiation or insolation is determined by $\delta
R_\text{a} = C_\text{a} F_1$; $\delta R_\text{o} = C_\text{o} F_1$. In
Eq.~(\ref{eq:psiodec}), a term $\overline{\phi_j}$ is added to the oceanic
basis function $\phi_j(x',y')$ in order to give it a vanishing spatial
average. This is required to guarantee mass conservation in the ocean
\citep{CP2001,McW1977}, but otherwise does not affect the dynamics. Indeed,
it can be added a posteriori when plotting the field
$\psi_\text{o}(x',y',t)$. This term is non-zero for odd $P_\text{o}$ and
$H_\text{o}$,
\begin{align}
 \overline{\phi_j} &= \frac{n}{2\pi^2} \int _0^{\pi }\int _0^{\frac{2 \pi }{n}}\phi_j(x',y') \text{d}x' \text{d}y'  \\
                   &= 2\frac{((-1)^{H_\text{o}} - 1) ((-1)^{P_\text{o}} - 1)}{H_\text{o} P_\text{o} \pi^2}.\nonumber
\end{align}
The mass conservation is automatically satisfied for $\psi_\text{a}
(x',y',t)$, as the spatial averages of the atmospheric basis functions
$F_i(x',y')$ are zero.

Substituting the fields in Eqs.~(\ref{eqn:atmos1})--(\ref{eqn:heat_ocean})
and projecting on the different basis functions yield
$2(n_\text{a}+n_\text{o})$ ordinary differential equations (ODEs) for as many
variables. Due to the linearization of the quartic temperature fields in
Eqs.~(\ref{eqn:heat_atmos}) and (\ref{eqn:heat_ocean}), these equations are
at most bilinear (due to the advection term) in the variables
$\psi_{\text{a},i} ,\theta_{\text{a},i}, \psi_{\text{o},j} $, and $\delta
T_{\text{o},j}$, which will henceforth jointly be referred to as $\eta_i$,
the components of the state vector $\vec{\eta}$.

To construct the dynamical equations of these variables, one has to compute
the various projections or inner products with the basis functions, for which
the following shorthand notation will be used:
\begin{equation}
\langle S,G \rangle \equiv \frac{n}{2\pi^2}\int_0^\pi\int_0^{2\pi/n}
S(x',y')\;G(x',y')\text{d}x'\; \text{d}y'.
\end{equation}

As described by \citet{CT1987}, the inner products for the atmosphere can be
computed as purely algebraic formulae of the wave numbers $P$, $M$, and $H$.
We reiterate these algebraic formulae in Sect.~\ref{sec:atmo-coef} of
Appendix~\ref{app:formulae} and extend them with the formulae for both the
ocean--atmosphere coupling terms and the ocean inner products in
Sect.~\ref{sec:oc-coef}. The inner products can be represented as either
two-dimensional or three-dimensional tensors, which are sparse but generally
not diagonal.

\subsection{Technical implementation}

Substituting the fields by Eqs.~(\ref{eq:psidec})--(\ref{eq:Tdec}) and
calculating the coefficients using the expressions for the inner products as
in Appendix~\ref{app:formulae} yields a set of $N \equiv
2(n_\text{a}+n_\text{o})$ prognostic ordinary differential equations. These
equations are at most bilinear in the variables $\eta_i$ $(1\leq i\leq N$)
due to the linearization of the radiative terms around a reference
temperature present in Eqs.~(\ref{eqn:heat_atmos})--(\ref{eqn:heat_ocean}).
This system of ODEs can therefore be most generically expressed as the sum of
a constant, a matrix multiplication, and a tensor contraction:
\begin{align}
\frac{\text{d}\eta_i}{\text{d}t} &= c_i + \sum_{j=1}^{N} m_{i,j} \; \eta_j  \label{eqn:orig_set} + \sum_{j,k=1}^{N} t_{i,j,k} \eta_j \eta_k \quad (1\leq i\leq
N).
\end{align}
This expression can be further simplified by adding a dummy variable that is
identically equal to one: $\eta_0\equiv 1$. This extra variable allows one to
merge $c_i$, $m_{i,j}$, and $t_{i,j,k}$ into the tensor
$\mathcal{T}_{i,j,k}$, in which the linear terms are represented by
$\mathcal{T}_{i,j,0}$ and the constant term by $\mathcal{T}_{i,0,0}$:
\begin{align}
&\frac{\text{d}\eta_i}{\text{d}t} = \sum_{j=0}^{N} \sum_{k=0}^{N}
\mathcal{T}_{i,j,k} \; \eta_j \; \eta_k \quad (1\leq i\leq N).
\label{eqn:tenscont}
\end{align}
The elements of the tensor $\mathcal{T}_{i,j,k}$ are specified in
Appendix~\ref{app:tensor}. Recasting the system of ordinary differential
equations for $\eta_i$ in the form of a tensor contraction has certain
advantages, as we will clarify below. The symmetry of
Eq.~(\ref{eqn:tenscont}) allows for a unique representation of
$\mathcal{T}_{i,j,k}$, if it is taken to be upper triangular in the last two
indices ($\mathcal{T}_{i,j,k} \equiv 0 $ if $j > k$). Since
$\mathcal{T}_{i,j,k}$ is known to be sparse, it is stored using the
coordinate list representation, i.e. a list of tuples
$(i,j,k,\mathcal{T}_{i,j,k})$. This representation renders the computation of
the tendencies $\text{d}\eta_i/\text{d}t$ computationally very efficient as
well as conveniently parallelizable.

Two implementations of \textsc{maooam} are provided as a Supplement: one in
Lua and one in Fortran. The Lua code is optimized for LuaJIT, a just-in-time
compiler for Lua \citep{luajit}, and runs about 20\,{\%} slower than the
Fortran version. By default, the model equations are numerically integrated
using the Heun method. We have tested higher-accuracy methods, but these did
not significantly change the results. The integration method can easily be
changed; as an example, a fourth-order Runge--Kutta integrator is also
included in the Lua implementation.

\subsection{Derivation of Jacobian, tangent linear, and adjoint models}

The form of Eq.~(\ref{eqn:tenscont}) allows one to easily compute the
Jacobian matrix of this system of ODEs. Indeed, denoting the right-hand side
of Eq.~(\ref{eqn:tenscont}) as $\text{d}\eta_i/\text{d}t = f_i$, the
expression reduces to
\begin{align}
  J_{i,j} = \frac{\text{d}f_i}{\text{d}\eta_j}& = \text{d} (\sum_{k,l=0}^N \mathcal{T}_{i,k,l} \; \eta_k \; \eta_l ) / \text{d}\eta_j  \\
                  &= \sum_{k=0}^N \left ( \mathcal{T}_{i,k,j} + \mathcal{T}_{i,j,k} \right) \eta_k \quad (1\leq i,j\leq N).\nonumber
\end{align}
The differential form of the tangent linear (TL) model for a small
perturbation $\boldsymbol{\delta\eta}^\text{TL}$ of a trajectory
$\boldsymbol{\eta}^{\ast}$ is then simply \citep{K2003}
\begin{align}
 \frac{\text{d}\delta\eta_i^\text{TL}}{\text{d}t} &= \sum_{j=1}^N J^{\ast}_{i,j} \; \delta\eta_j^\text{TL}   \\
             &= \sum_{j=1}^N\sum_{k=0}^N \left ( \mathcal{T}_{i,k,j} + \mathcal{T}_{i,j,k} \right) \eta^{\ast}_k \; \delta\eta_j^\text{TL}\quad (1\leq i\leq N).\nonumber
\end{align}
To obtain the differential form of the adjoint model along the trajectory
$\boldsymbol{\eta}^{\ast}$, the Jacobian is transposed to yield the following
equations for the adjoint variable $\boldsymbol{\delta\eta}^\text{AD}$:
\begin{align}
 -\frac{\text{d}\delta\eta^\text{AD}_i}{\text{d}t} &= \sum_{j=1}^N J^{\ast}_{j,i} \; \delta\eta^\text{AD}_j  \\
             &= \sum_{j=1}^N \sum_{k=0}^N \left ( \mathcal{T}_{j,k,i} + \mathcal{T}_{j,i,k} \right) \eta^{\ast}_k \; \delta\eta^\text{AD}_j \quad (1\leq i\leq N). \nonumber
\end{align}

\section{Model dynamics}\label{sec:dynamics}

This section details some key results obtained with the model for various
levels of spectral truncation, with the set of parameter values given in
Table~\ref{tab:params}. The parameter values for $L$, $L_\text{R}$,
$\lambda$, $r$, $d$, $C_\text{o}$, $C_\text{a}$, $k_d$, and $k_d^{\prime}$
were selected as detailed in \citet{VDDG2015}. The same value was chosen for
$k_d$ and $k_d^{\prime}$, as was done in \citet{CS1980}; see also
\citet{VD2014}. Unless otherwise stated, all the following results are
obtained after first integrating the model for a transient period of
30\,726.5~years. The model is subsequently integrated for another
92\,179.6~years to obtain a sufficiently long trajectory from which good
statistics can be extracted.

For the atmospheric part of the model, a previous study \citep{CT1987},
referred to as CT in the following, has shown that spurious chaos and a too
large variability in the modes near the spectral cut-off could take place if
the resolution is not high enough. These manifestations of spurious behaviour
can lead to solutions that differ significantly from the solutions of the
full partial differential equations (PDEs, here
Eqs.~\ref{eqn:atmos1}--\ref{eqn:heat_ocean}). These findings lead us to the
important question of \textit{convergence}: to what degree has the solution
of the truncated equations converged towards the solution of the PDEs?
Although we do not have access to the latter, one can infer how the solutions
are altered when the resolution is increased. Therefore, it cannot be
asserted that convergence has been reached, and this point was also clearly
stated in CT. However, we can reasonably suppose that when the solutions
stabilize, they give an insight into the full dynamics.

\begin{figure}[t]
\includegraphics[width=0.9\textwidth]{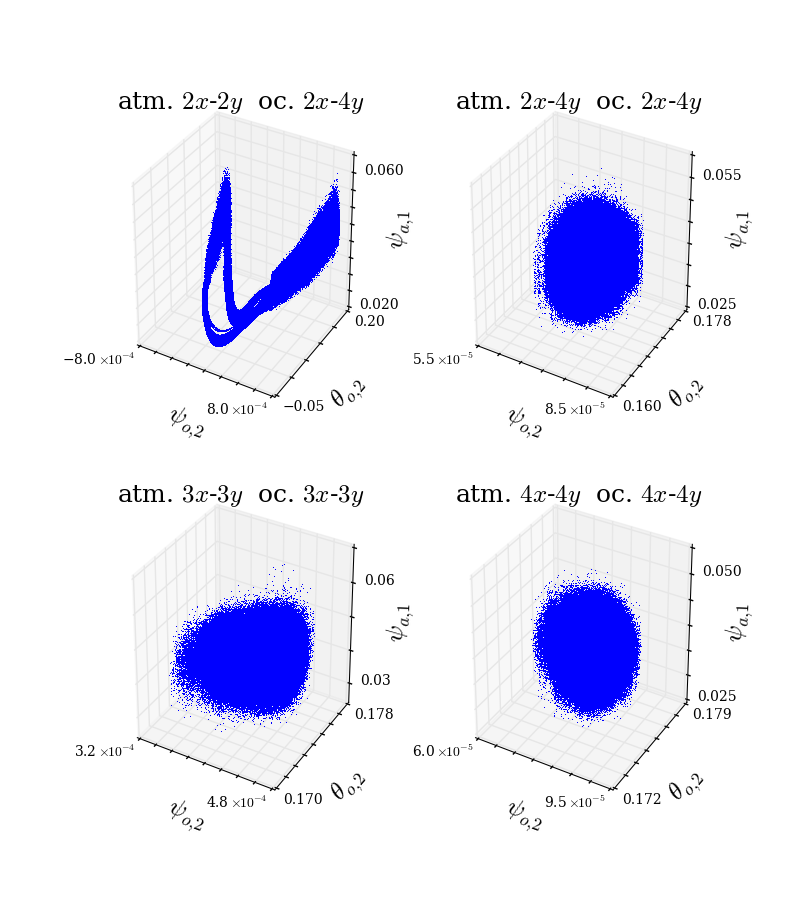}
\caption{Cross section of the attractors for various model resolutions. The
atmospheric and oceanic resolutions are both indicated above each panel. The
parameters are given in Table~\ref{tab:params}.}\label{fig:attractors_GRL2}
\end{figure}

\begin{figure}[t]
\includegraphics[width=0.9\textwidth]{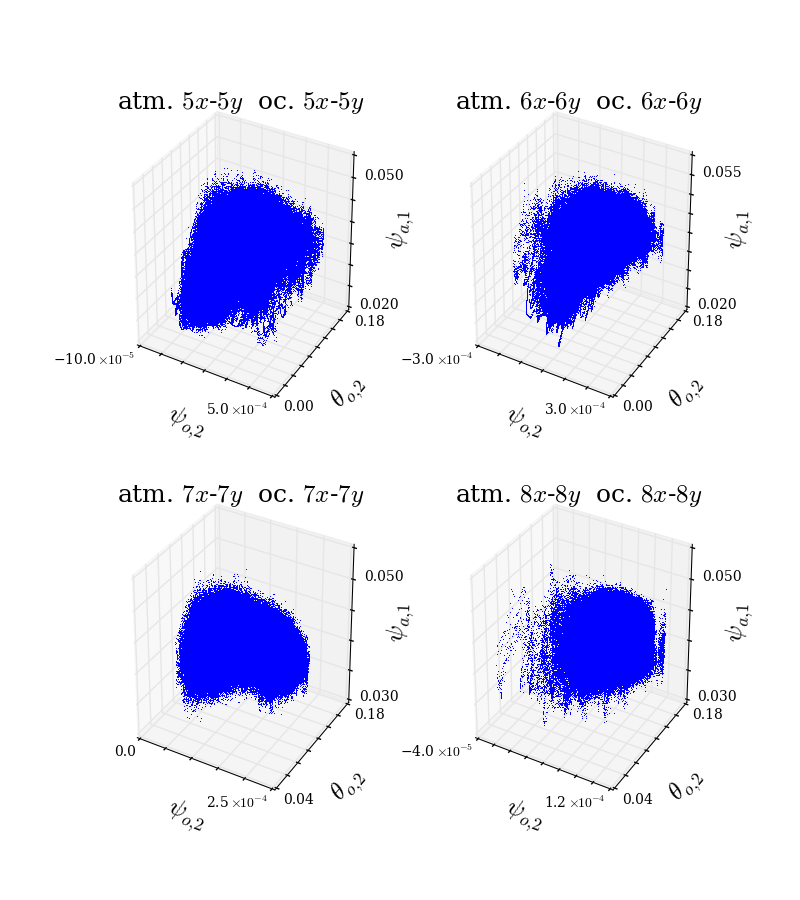}
\caption{Cross section of the attractors for various model resolutions
(continued from Fig.~\ref{fig:attractors_GRL2}).
\hack{\vspace*{9mm}}}\label{fig:attractors_GRL2_2}
\end{figure}

This question is now addressed for the \textsc{maooam} coupled
atmosphere--ocean model. Figures~\ref{fig:attractors_GRL2} and
\ref{fig:attractors_GRL2_2} display cross sections of the attractors of the
model for different resolutions. The three variables selected in this
projection are $\psi_{\text{a},1}$, $\psi_{\text{o},2}$, and
$\theta_{\text{o},2}$, which have already been used to represent the
large-scale variability of the model~\citep{VDDG2015}. We use the same
notation as in CT to specify the resolution of each component: $(H^{\max})
x$--$(P^{\max}) y$ for the atmosphere and $(H^{\max}_\text{o})
x$--$(P^{\max}_\text{o}) y$ for the ocean, with $M^{\max}=H^{\max}$. All the
model configurations used are listed in Table~\ref{tab:runtime}. To alleviate
the notation in the following, a model configuration denoted simply by
$H^{\max}x$--$P^{\max}y$ indicates that the resolution is the same in both
components: $H^{\max}_\text{o}=H^{\max}$ and $P^{\max}_\text{o}=P^{\max}$.

The first panel of Fig.~\ref{fig:attractors_GRL2}, with the atm. $2x$--$2y$
oc. $2x$--$4y$ resolution, shows the typical attractor geometry found
in~\citet{VDDG2015} and~\citet{V2015} with a noisy, seemingly periodic orbit
associated with the development of a large low-frequency signal. However, as
the resolution is increased in both the ocean and atmosphere components, this
structure destabilizes and we obtain more compact, noisy attractors in
Figs.~\ref{fig:attractors_GRL2} and \ref{fig:attractors_GRL2_2}. The cause of
this structural change is an interesting question in itself, which is worth
exploring further in the future, as it is associated with the problem of
structural stability of models, but is beyond the scope of the present work.

Regarding the question of convergence, the variability of the atmospheric
variables becomes quite stable as the resolution increases beyond $6x$--$6y$.
Indeed, the bounds of the attractors on the vertical axis
($\psi_{\text{a},1}$) stabilize at this resolution. This result is in
agreement with the findings of CT. On the other hand, the convergence is not
yet reached for the oceanic variables whose variability is strongly affected
by adding further modes as in the $7x$--$7y$ and $8x$--$8y$ resolutions.

\begin{figure}[t]
\includegraphics[width=0.6\textwidth]{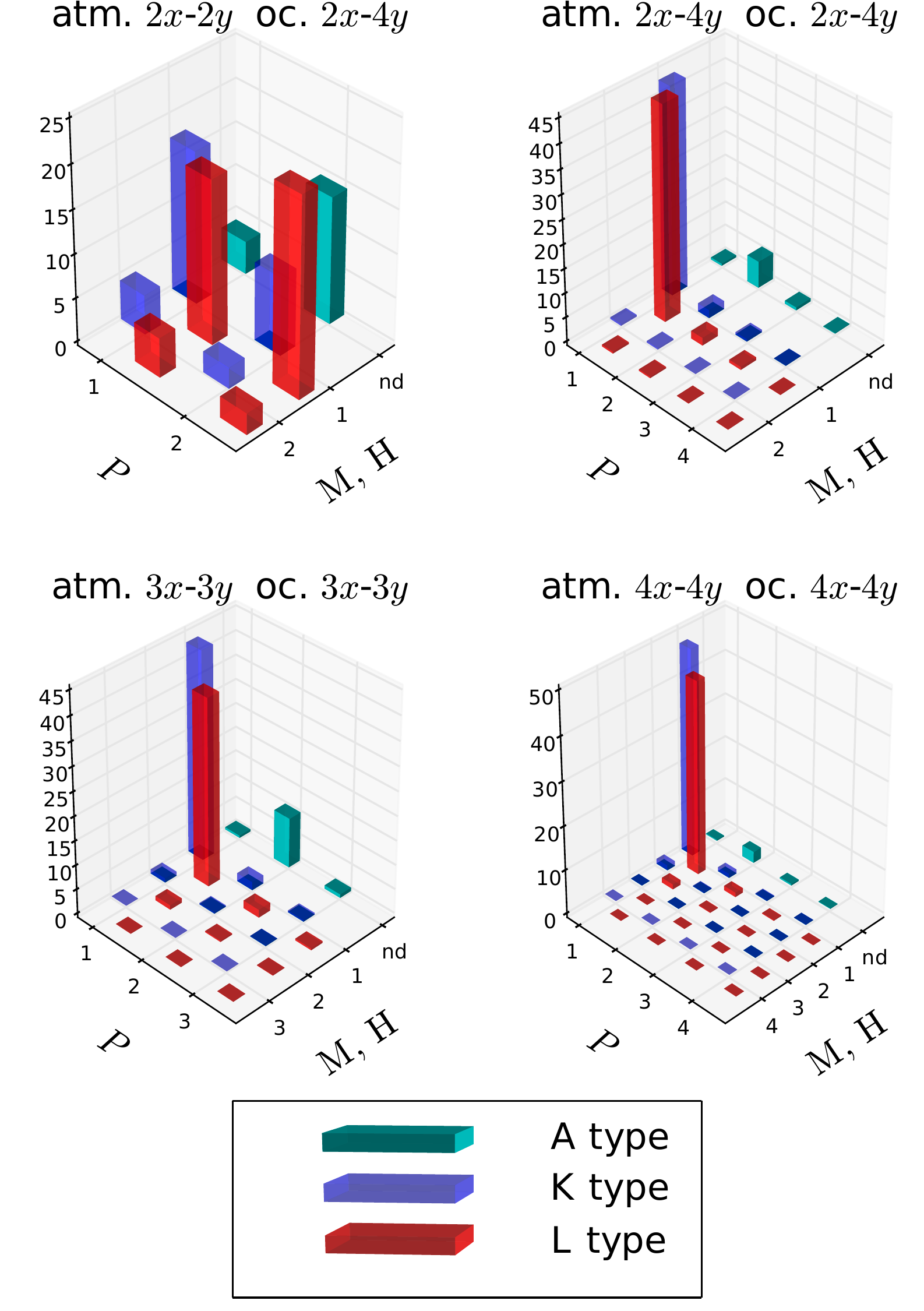}
\caption{Variance distributions of the $\psi_{\text{a},i}$ variables in
percents for various model resolutions. For the variables associated with the
$A$-type basis functions, the wave numbers $M$ and $H$ are not defined (nd).
\label{fig:var_psi_GRL2}}
\end{figure}

\begin{figure}[t]
\includegraphics[width=0.6\textwidth]{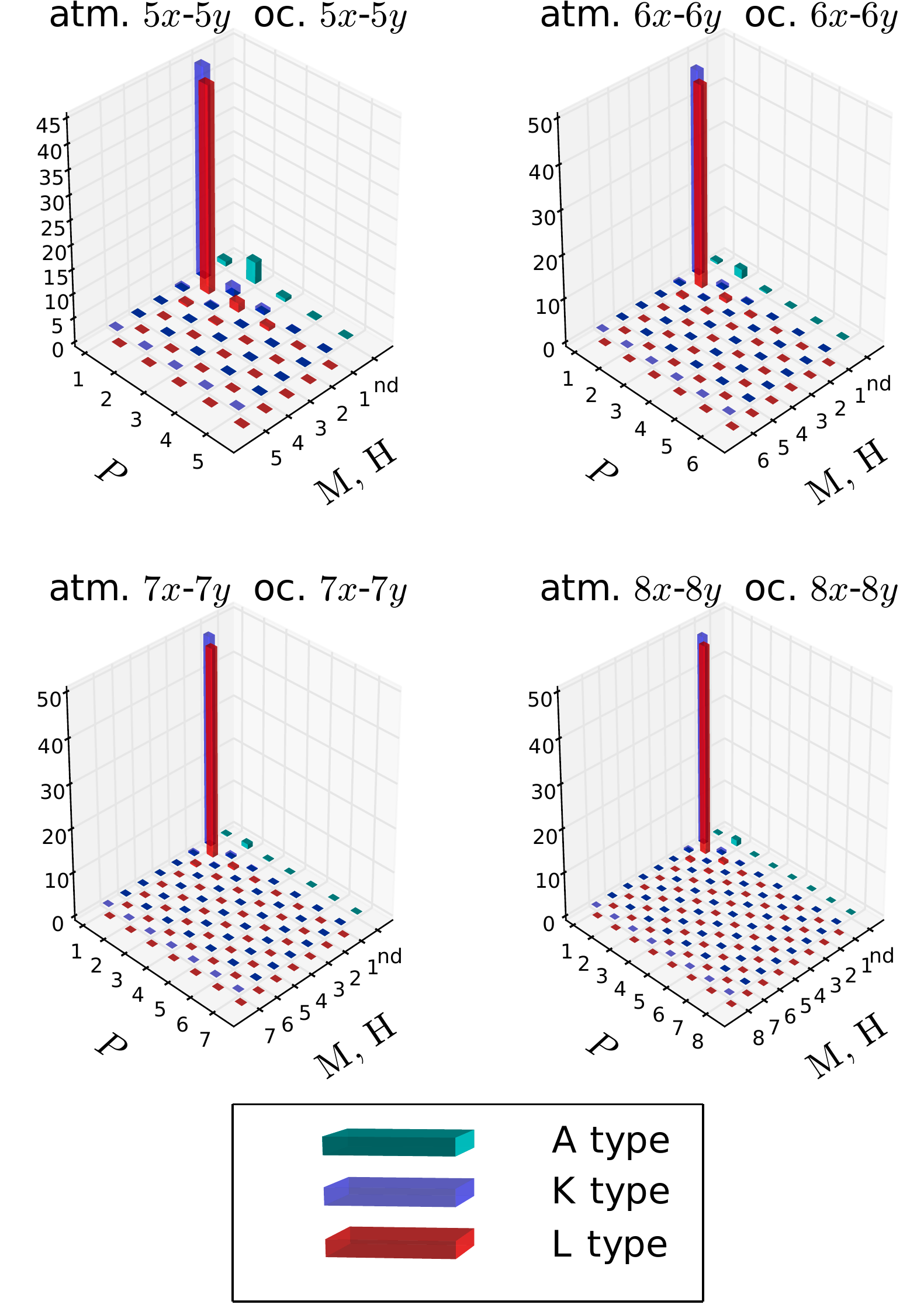}
\caption{Variance distributions of the $\psi_{\text{a},i}$ variables in
percents for various model resolutions (continued from
Fig.~\ref{fig:var_psi_GRL2}). For the variables associated with the $A$-type
basis functions, the wave numbers $M$ and $H$ are not defined (nd).
\label{fig:var_psi_GRL2_2}}
\end{figure}

\begin{figure}[t]
\includegraphics[width=0.9\textwidth]{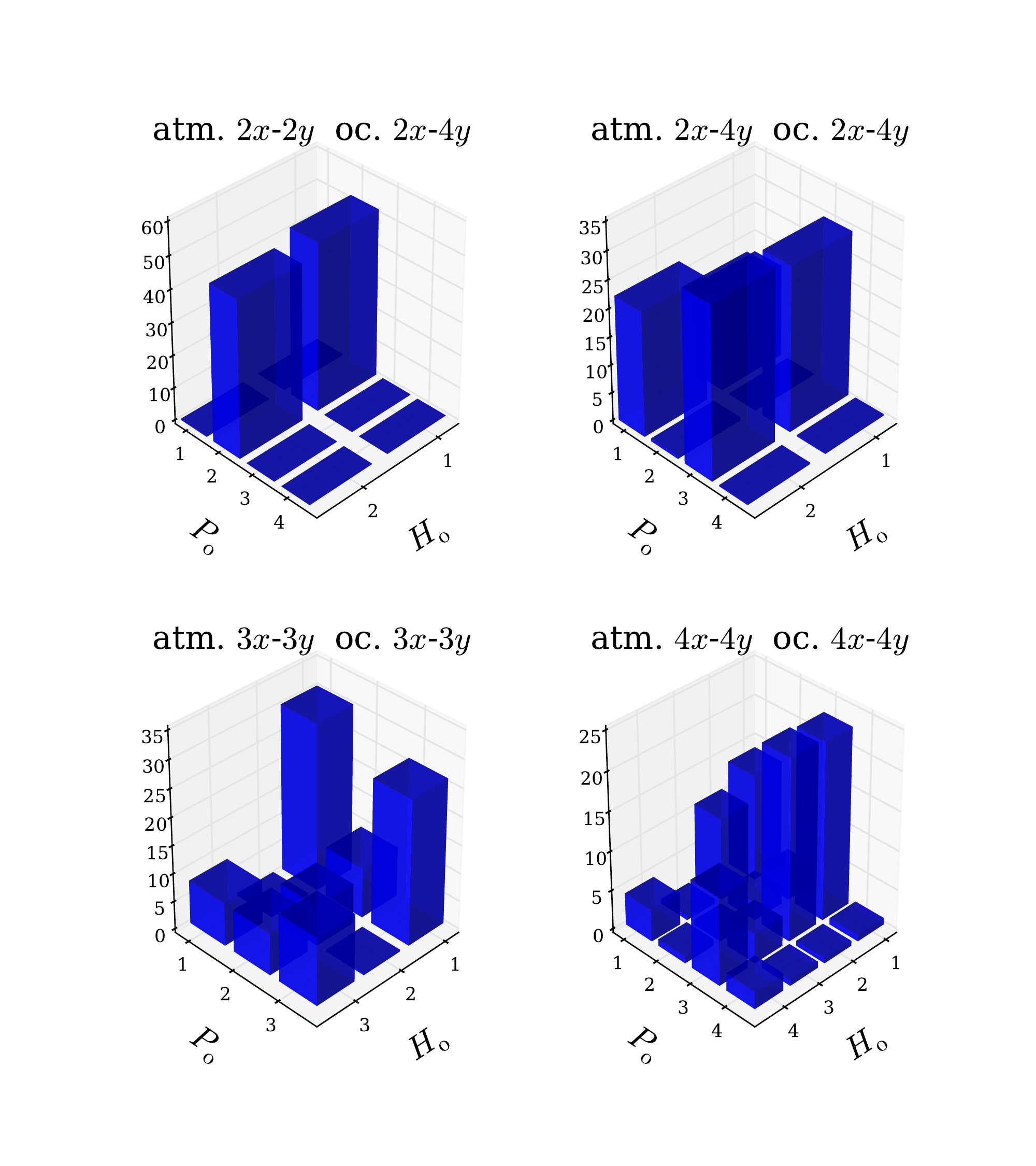}
\caption{Variance distributions of the $\psi_{\text{o},i}$ variables in
percents for various model resolutions. \label{fig:var_A_GRL2}}
\end{figure}

\begin{figure}[t]
\includegraphics[width=0.9\textwidth]{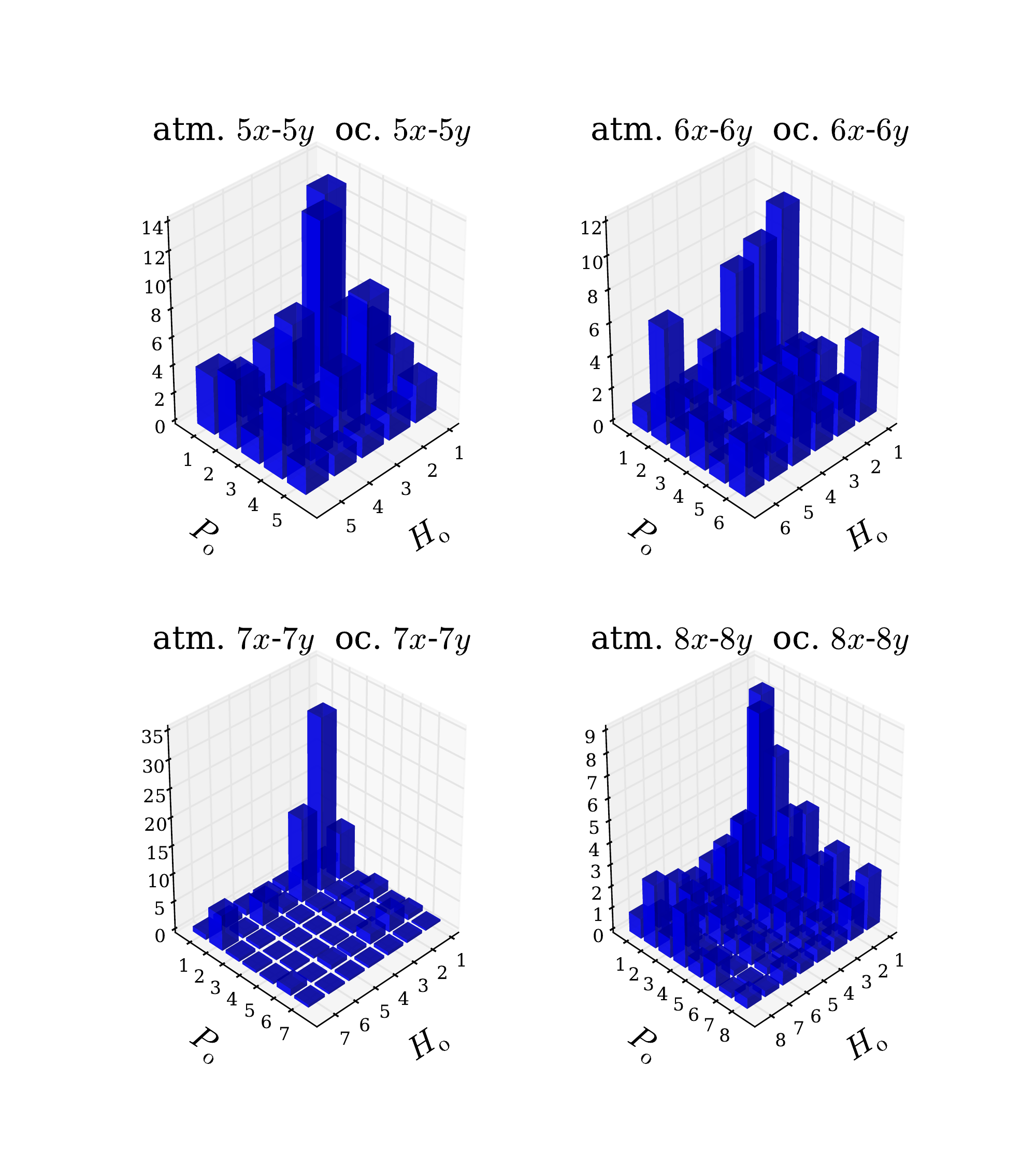}
\caption{Variance distributions of the $\psi_{\text{o},i}$ variables in
percents for various model resolutions (continued from
Fig.~\ref{fig:var_A_GRL2}). \label{fig:var_A_GRL2_2}}
\end{figure}

The impact of the resolution on the solutions can also be examined by
computing the variance of each variable of the barotropic and baroclinic
streamfunctions, since these are associated with the kinetic and potential
energy of the system \citep{Y1980}. The presence of spurious behaviour can
then be detected through substantial changes in this variability. The
distributions of the total variance of the variables $\psi_{\text{a},i}$ and
$\psi_{\text{o},i}$ are depicted in
Figs.~\ref{fig:var_psi_GRL2}--\ref{fig:var_A_GRL2_2}. The results show that
the variance distribution does not change much beyond the $4x$--$4y$
resolution for the atmospheric component. However, for the oceanic component,
the variance distribution is strongly modified when the resolution increases,
and therefore one cannot conclude from Fig.~\ref{fig:var_A_GRL2_2} that some
sort of convergence is reached at the $8x$--$8y$ resolution. To interpret
this specific property, one must recall an important feature of
two-dimensional quasi-geostrophic turbulence, namely the presence of a
specific space scale, the \textit{Rhines scale}, which delimits the two
regimes associated with a wave-dominated dynamics and a turbulent dynamics.
This space scale is given by

\begin{equation}
L_\textit{Rh}= \sqrt{\frac{U}{\beta}},
\end{equation}
where $U$ represents the root-mean-square velocity of the energy-containing
scales~\citep{R1975,VM1993,V2006} and $\beta=\text{d}f/\text{d}y$ is the
meridional derivative of the Coriolis parameter $f$. If one takes the typical
velocity of the order of a few metres per second and a few centimetres per
second within the atmosphere and the ocean at large scales, the typical
length scales will be of the order of 1000 and 100\,km, respectively.
Therefore the highest wave numbers necessary to resolve the wave-dominated
part within the atmosphere and the ocean differ by a factor of 10. Coming
back to our analysis, if this limit is reached for the atmosphere in our
model at $H$, $P=4$--5, we should suspect that a value of $H_\text{o}/2
\approx P_\text{o} \approx 40$--50 should be used for the ocean. This of
course imposes strong constraints on our reduced-order model and would
considerably limit its flexibility.

\begin{figure}[t]
\includegraphics[width=0.9\textwidth]{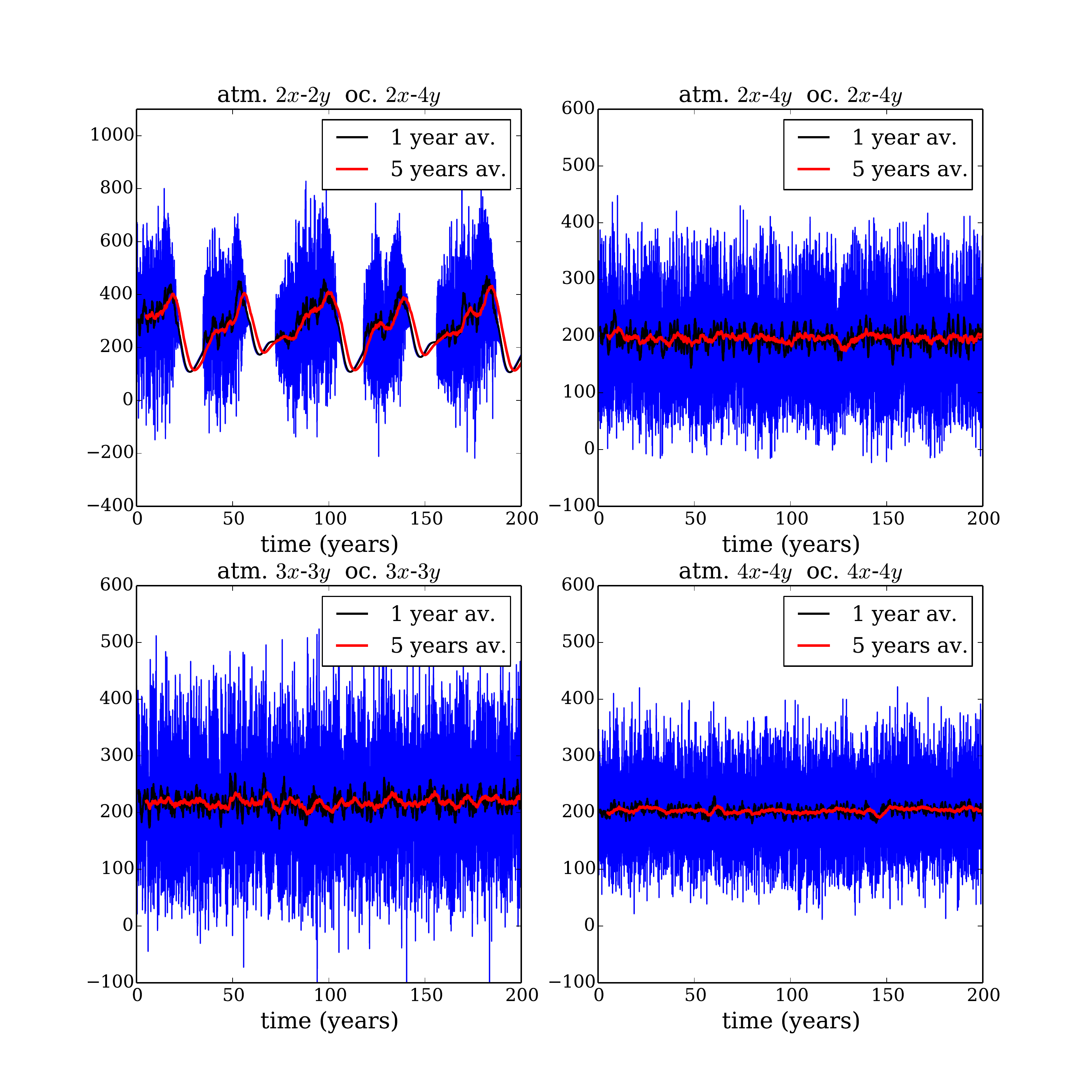}
\caption{Time series of the geopotential height difference (m) between
locations ($\pi/n, \pi/4)$ and ($\pi/n, 3 \pi/4)$ of the model's
non-dimensional domain for different resolutions. Running averages for
$\tau=1 y$ (black) and $\tau=5 y$ (red) are also provided, highlighting the
LFV signal present in the series.}\label{fig:LFV_GRL2}
\end{figure}

\begin{figure}[t]
\includegraphics[width=0.9\textwidth]{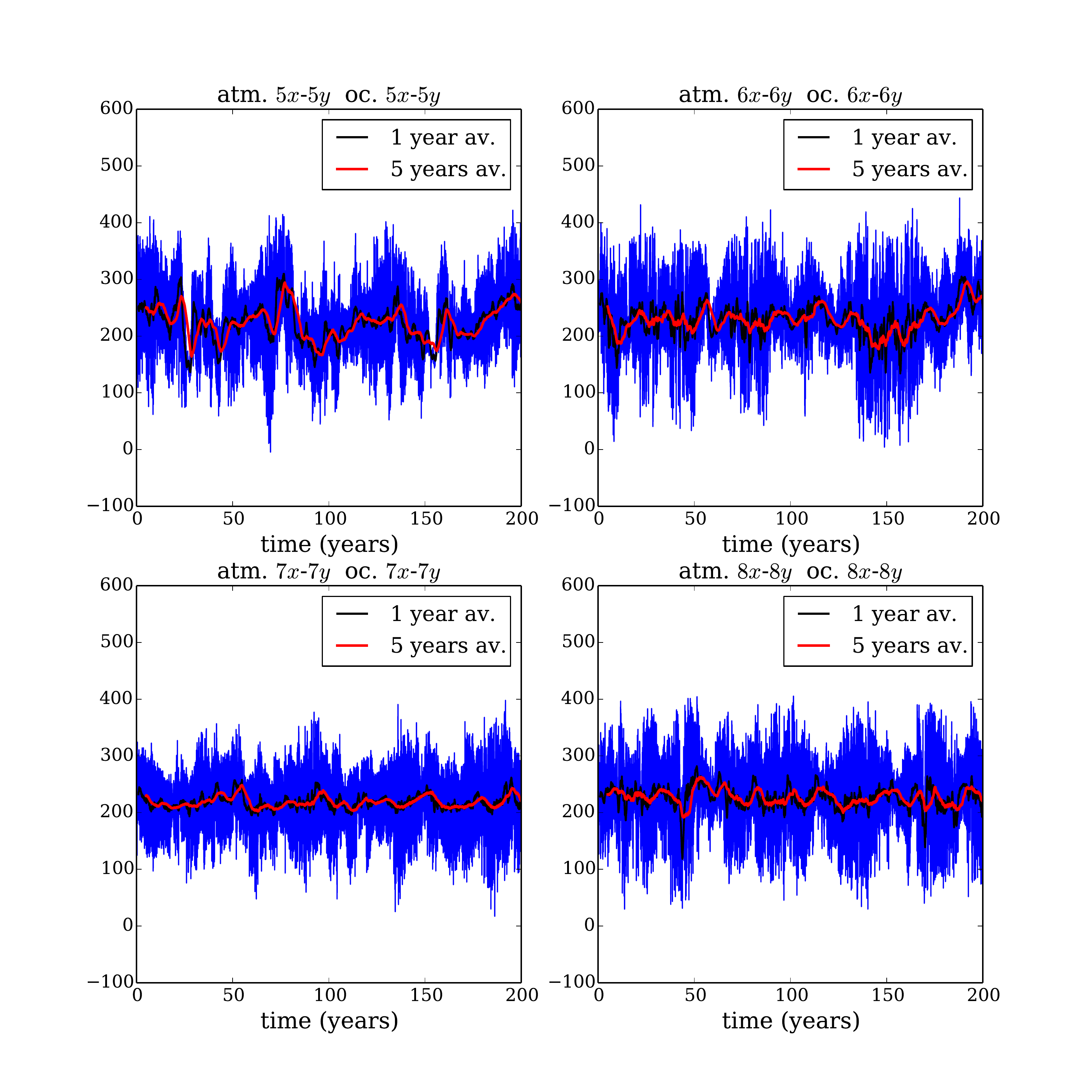}
\caption{Time series of the geopotential height difference (continued from
Fig.~\ref{fig:LFV_GRL2}).\hack{\vspace*{8mm}}}\label{fig:LFV_GRL2_2}
\end{figure}

Let us now focus on the development of the LFV in these different model
configurations, and let us define the geopotential height difference $\delta
z$ between the locations ($\pi/n, \pi/4$) and ($\pi/n, 3 \pi/4$) of the
model's non-dimensional domain:
\begin{align*}
&\delta z(t) = z(\pi/n, \pi/4,t) - z(\pi/n, 3 \pi/4,t),\\
&z(x^\prime,y^\prime,t) = \frac{f_0}{g} \,
\psi_\text{a}(x^\prime,y^\prime,t),
\end{align*}
where $z$ is the geopotential height field, as in \citet{VDDG2015}. The
results shown in Figs.~\ref{fig:LFV_GRL2} and \ref{fig:LFV_GRL2_2} indicate
that the LFV, present for atm. $2x$--$2y$ oc. $2x$--$4y$ as in \citet{V2015},
is a very weak signal at intermediate resolutions, but develops again when
the number of modes is increased, as shown by the 1-year and 5-year running
means. It suggests that the LFV previously found in low-resolution versions
(see Fig.~\ref{fig:LFV_GRL2}, panel atm. $2x$--$2y$ oc. $2x$--$4y$) is a
robust feature of the model. Moreover, at high resolutions this LFV is weaker
than for the VDDG model version, but it seems closer to the actual dynamics
found for the North Atlantic Oscillation (NAO) as discussed in \citet{LW2003}
and \citet{S2000}.

\begin{figure}[t]
\includegraphics[width=0.9\textwidth]{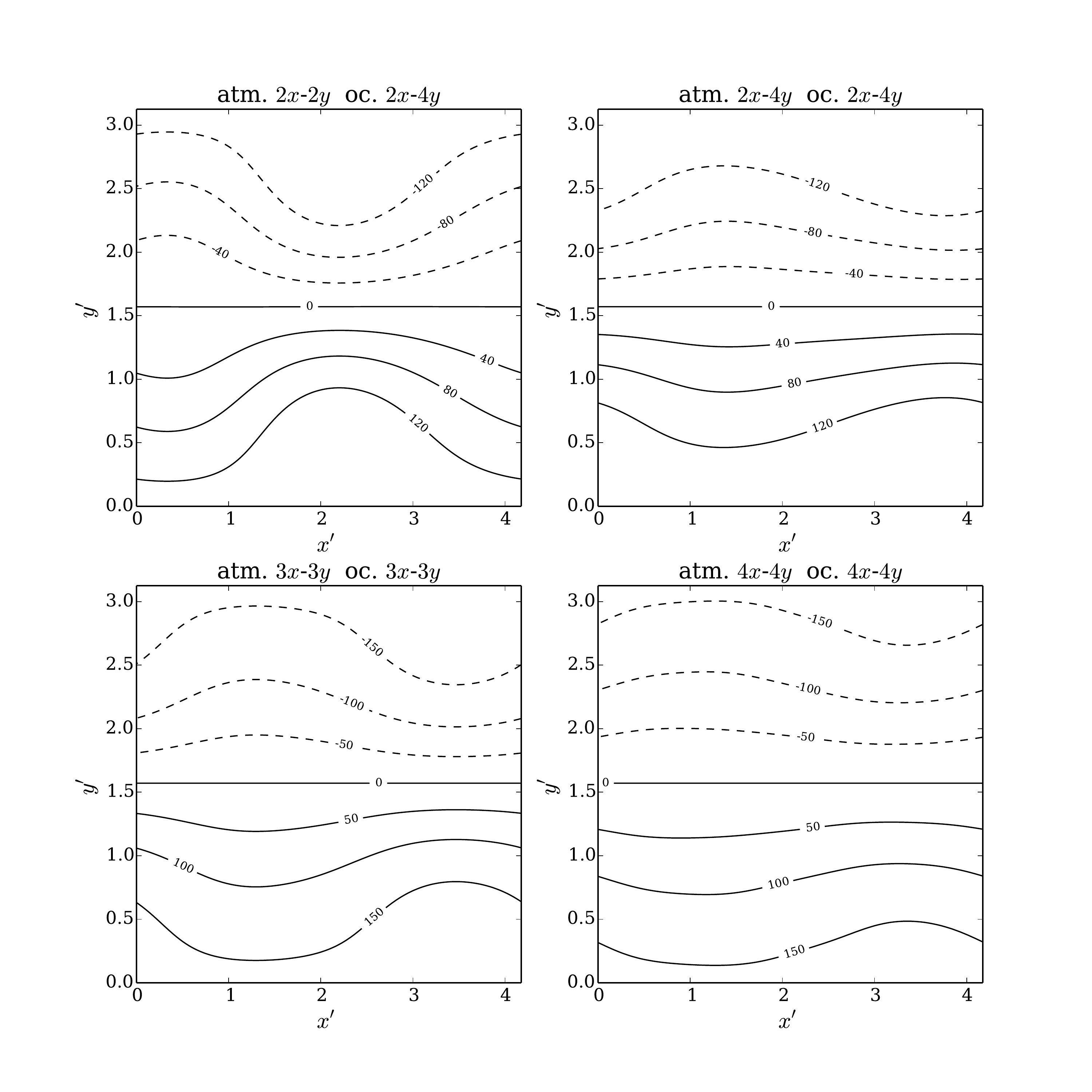}
\caption{Climatologies for the geopotential height field $z = \frac{f_0}{g}
 \psi_\text{a}$ (m) presented on the non-dimensional model domain, as
obtained using 92\,179.6~years of model integration.
\label{fig:clim_psi_GRL2}}
\end{figure}

\begin{figure}[t]
\includegraphics[width=0.9\textwidth]{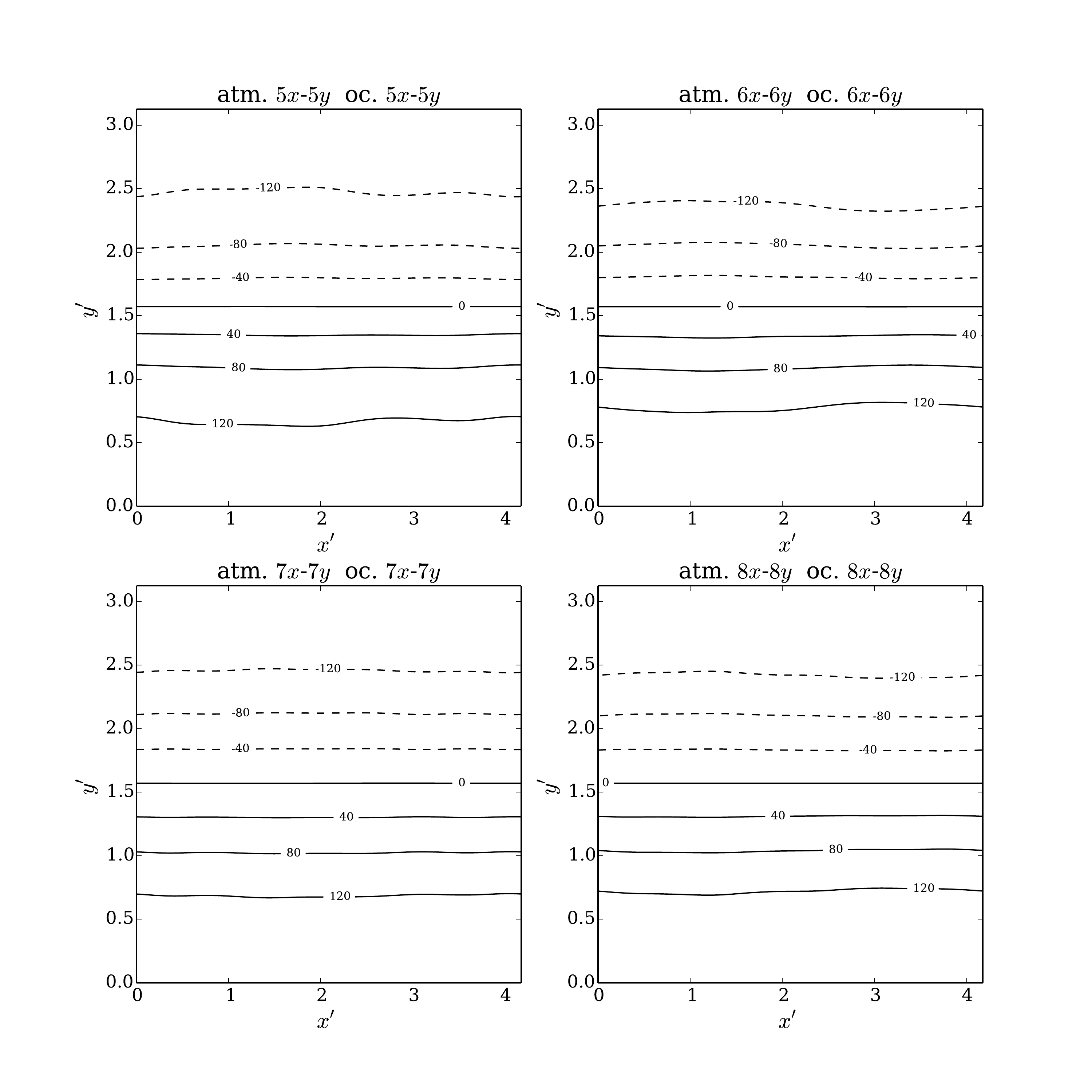}
\caption{Climatologies for the geopotential height field $z = \frac{f_0}{g}
 \psi_\text{a}$ (m) (continued from Fig.~\ref{fig:clim_psi_GRL2}).
\label{fig:clim_psi_GRL2_2}}
\end{figure}

\begin{figure}[t]
\includegraphics[width=0.9\textwidth]{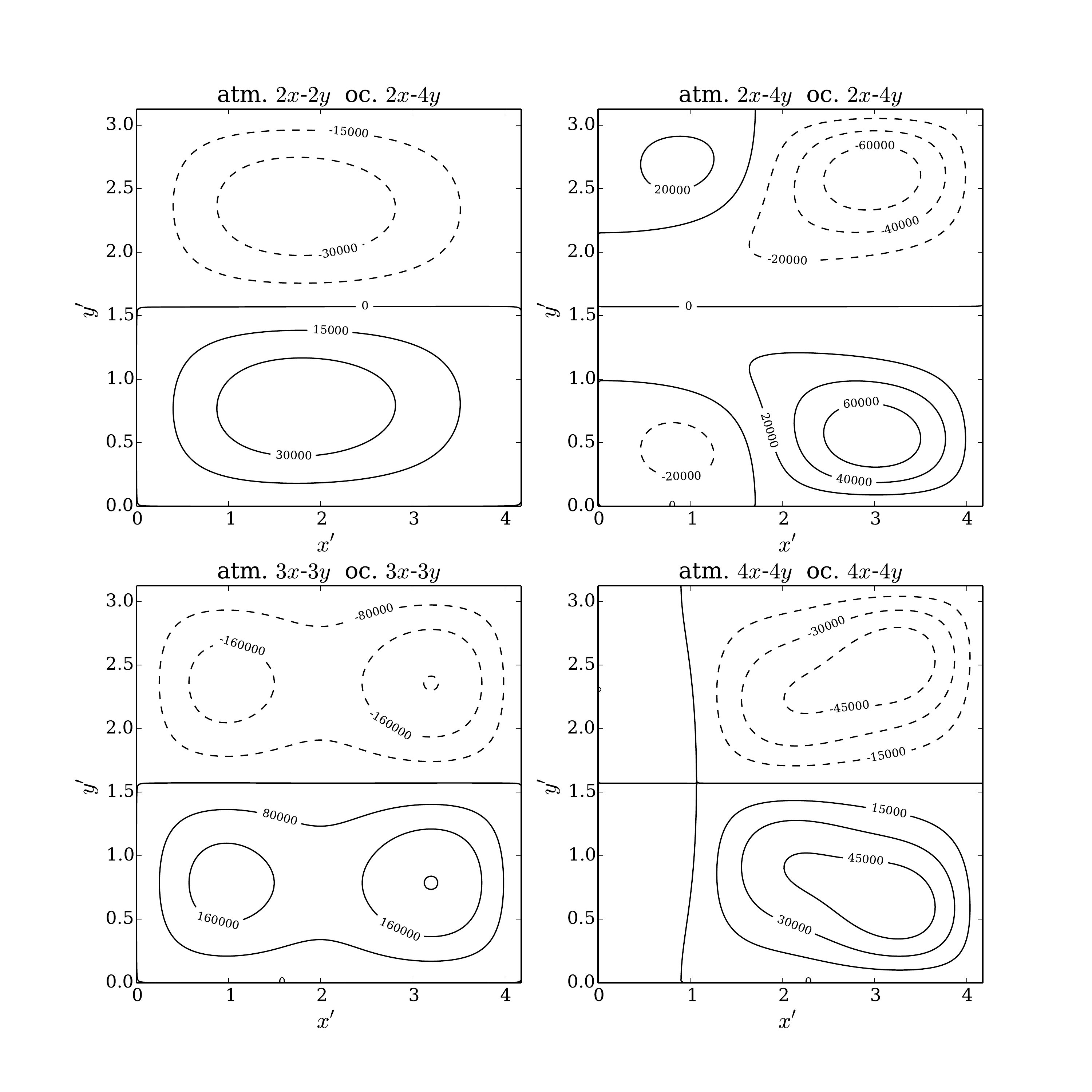}
\caption{Climatologies for the oceanic streamfunction field $\psi_\text{o}$
(m$^2$\,s$^{-1}$) presented on the non-dimensional model domain, as obtained
using 92\,179.6~years of model integration. \label{fig:clim_A_GRL2}}
\end{figure}

\begin{figure}[t]
\includegraphics[width=0.9\textwidth]{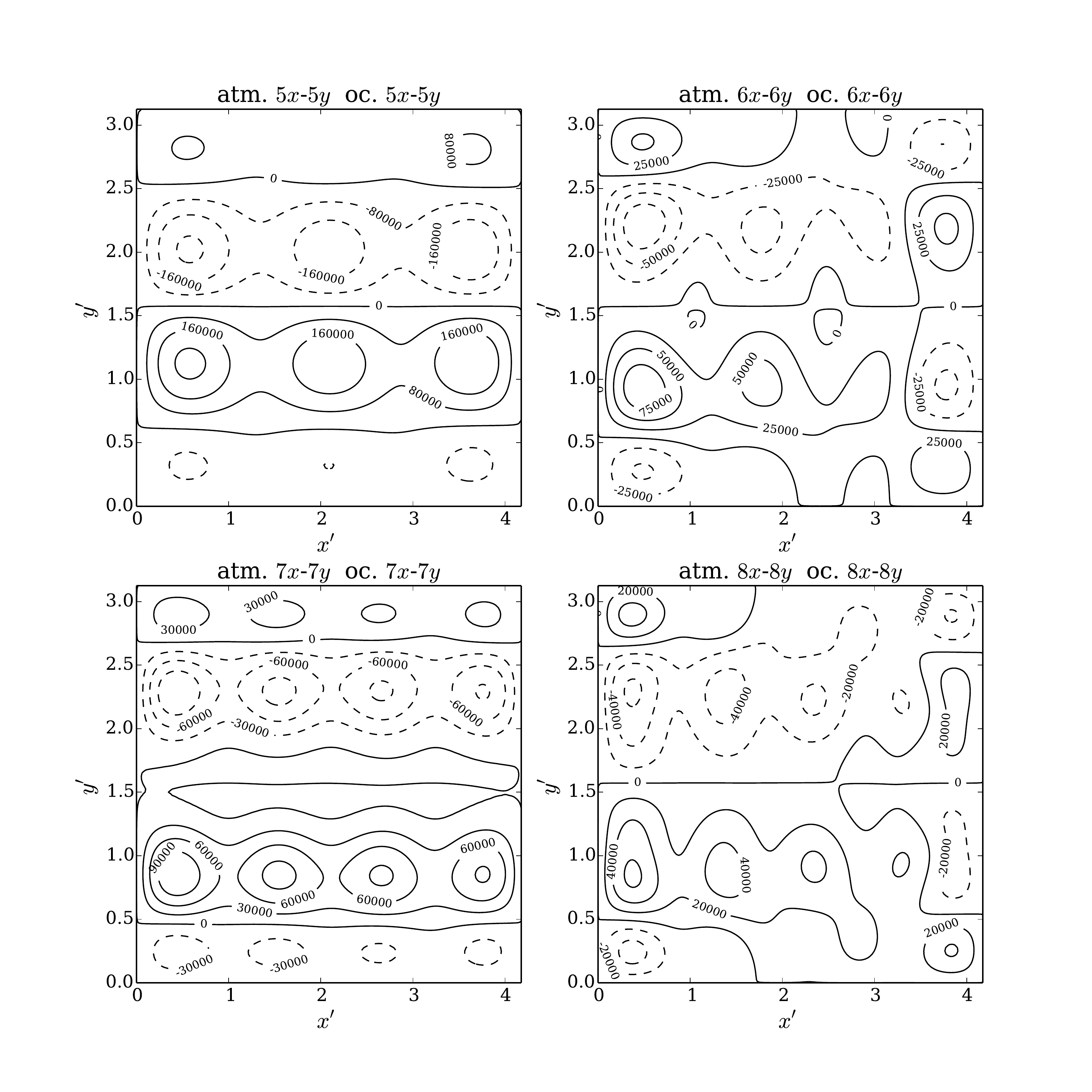}
\caption{Climatologies for the oceanic streamfunction field $\psi_\text{o}$
(m$^2$\,s$^{-1}$) (continued from Fig.~\ref{fig:clim_A_GRL2}).
\label{fig:clim_A_GRL2_2}}
\end{figure}

The climatologies of the atmospheric barotropic streamfunction expressed in
geopotential height further highlight the changes in the statistical
properties of the model as a function of resolution. As shown in
Figs.~\ref{fig:clim_psi_GRL2} and \ref{fig:clim_psi_GRL2_2}, the convergence
is pretty fast toward an averaged zonal atmospheric circulation as the model
resolution is increased. By contrast, the convergence for the oceanic
streamfunction $\psi_\text{o}$ is less clear (Figs.~\ref{fig:clim_A_GRL2} and
\ref{fig:clim_A_GRL2_2}), although a recurrent ``global'' double gyre is
present for each resolution. As for the LFV, the topology of the gyres at
high resolutions and their small-scale structures also seem to depend on
whether $H^{\max}$, $M^{\max}$, $H_\text{o}^{\max}$ and $P_\text{o}^{\max}$,
$P^{\max}$ are even or odd numbers.

The previous results point toward the important question of the optimal
resolution of the oceanic component needed to get a sufficiently
low-resolution model while keeping a dynamics with strong similarities to a
very high-resolution model. To answer this question, we have performed some
higher-resolution integrations, but on shorter time spans. The time span for
each integration is given in Table~\ref{tab:runtime}.

\begin{figure}[t]
\includegraphics[width=0.9\textwidth]{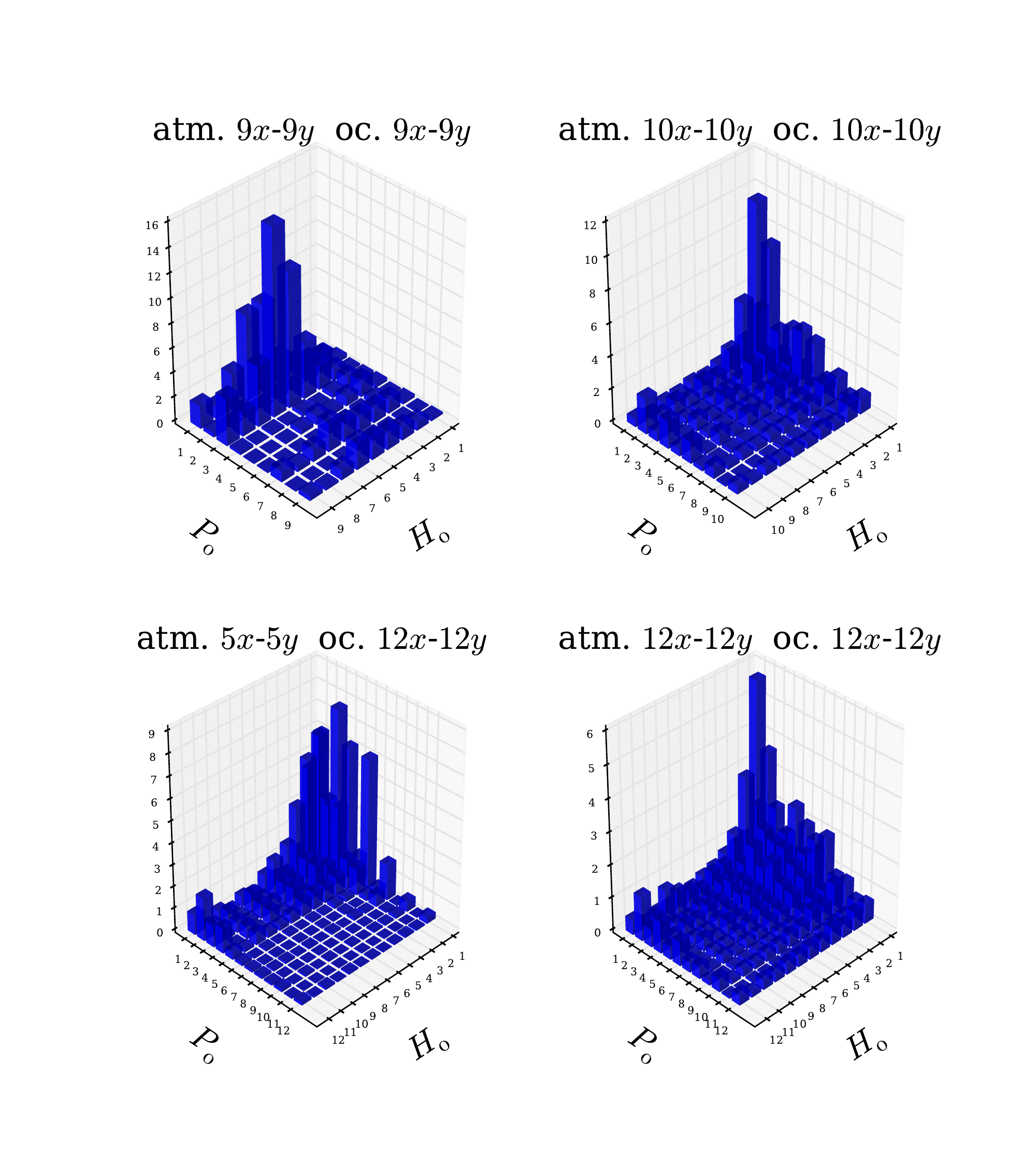}
\caption{Variance distributions of the $\psi_{\text{o},i}$ variables in
percents for the high-resolution runs. The information on the different
runtimes is gathered in Table~\ref{tab:runtime}. \label{fig:var_A_GRL2_high}}
\end{figure}

The variance distributions of the oceanic streamfunction variables (see
Fig.~\ref{fig:var_A_GRL2_high}) have decreased at the spectral cut-off's
edges compared to the distributions of the lower-resolution model
configurations shown in Fig.~\ref{fig:var_A_GRL2}. However, this decrease is
not sufficient, and apparently spurious effects are still present. For
instance, the decay is not identical in both directions, with a slower decay
rate as the zonal wave number $H_\text{o}$ increases. We can even notice a
peak in the distribution around $H_0=H_0^{\max}$ and $P_0=2$ for all these
higher model resolutions. This indicates that in fact we are still far from a
quantitatively representative solution in the ocean. It confirms that, as
stated previously, a resolution of the order of the Rhines scale is needed to
achieve a good convergence. For the ocean, it corresponds to a 100\,km
resolution which would then require roughly 2000 modes. Such a model will of
course be very computationally expensive and cannot be considered a
``reduced''-order model anymore.

However, the comparison between the atm. $5x$--$5y$ oc. $12x$--$12y$ model
configuration and the $10x$--$10y$ or $12x$--$12y$ model configurations shows
that the former displays a large-scale behaviour close to the latter two, but
with a reduced complexity and computational cost. This similarity can be
assessed by considering the climatologies of these higher-resolution runs
displayed in Fig.~\ref{fig:clim_A_GRL2_high} and by watching the
corresponding videos (see below). We therefore believe that the atm.
$5x$--$5y$ oc. $12x$--$12y$ model configuration is a good candidate when
investigating more realistic dynamics than the one presented in
\textsc{vddg}. It must however be stressed that the \textsc{vddg} model is
still an important tool in this hierarchy of models since it already contains
the basic mechanisms leading to low-frequency variability. In addition, the
climatologies shown in Fig.~\ref{fig:clim_A_GRL2_high} confirm the dependence
of the dynamics on whether $H^{\max}$, $M^{\max}$, $H_\text{o}^{\max}$ and
$P_\text{o}^{\max}$, $P^{\max}$ are even or odd, and also the presence of a
global double-gyre in the ocean.

Finally, the dynamics of the model for the various resolutions are also
illustrated in the videos provided as supplementary material. These videos
depict the time evolution of the streamfunction and temperature fields, as
well as the geopotential height difference and the three-dimensional
phase-space projection shown in Figs.~\ref{fig:attractors_GRL2} and
\ref{fig:attractors_GRL2_2}. They give an insight into the high-frequency
atmospheric and low-frequency oceanic variability, and also show the
interesting time evolution of the oceanic gyres. In these videos, a striking
feature is the presence of a westward wave propagation within the ocean while
the LFV is developing in the coupled system. This feature has been associated
with the propagation of Rossby-like waves~\citep{V2015}.

\conclusions
\label{sec:conclusions}

A new reduced-order coupled ocean--atmosphere model is presented, extending
the low-resolution versions previously published \citep{VD2014,VDDG2015}. It
is referred to as \textsc{maooam}, for Modular Arbitrary-Order
Ocean-Atmosphere Model. This new model retains the main features of the
previous versions but allows for the selection of an arbitrary resolution
within the ocean and the atmosphere. Besides the potential utility of this
new functionality for evaluating the impact of the number of modes on the
dynamics (as has been done here), it opens the possibility of addressing
several new questions in a very flexible way, such as the development of a
consistent stochastic parameterization scheme through scales, the
understanding of the predictability problem at multiple scales and the role
of model error, or the implementation of a data-assimilation scheme for the
coupled ocean--atmosphere system.

In the present work, we have studied the impact of the resolution on the
model solution's dynamics, by investigating the properties of the attractors
and the variance distributions in both the oceanic and atmospheric
components. The conclusion that can be drawn is that the convergence of the
atmospheric component of the system is quite fast (as noted in
\citealp{CT1987}), with variance distributions decreasing rapidly as a
function of scale. However, the convergence of the oceanic component is much
slower. Consequently, none of the solutions presented so far have
satisfactorily converged toward a dynamics that correctly reflects the
wave-dominated regime of the coupled ocean--atmosphere system. This regime
corresponds to a resolution associated with the Rhines scale (which for the
ocean is equal to 100\,km or, equivalently, to wave numbers of the order of
$H_\text{o}^{\max}/2 \approx P_\text{o}^{\max} \approx 50$). This stresses
the need for high-resolution oceanic models to correctly represent the full
coupled dynamics. One coupled model configuration which could, however, be
recommended so far is the atm. $5x$--$5y$ oc. $12x$--$12y$ configuration,
which seems to display some robustness in the ocean climatology as compared
to the full $10x$--$10y$ and $12x$--$12y$ configurations. This conclusion
requires further investigation with even higher resolutions, together with
the use of more advanced tools of analysis like the computation of the
Lyapunov exponents as in~\citet{VL2016}. These can be computed using the
tangent linear model version for which an implementation is also provided.
This will be the subject of a future investigation.

The robustness of the LFV pattern, one of the most interesting features of
the model, has also been explored. As it turns out, a LFV is still present in
a large portion of the model configurations explored (not in $2x$--$4y$,
$3x$--$3y$, and $4x$--$4y$), but a weaker LFV signal is found when
high-resolution configurations are used. A dominant signal is found with a
wide variety of periods ranging from 1 to 100~years, depending on the model
configuration. A more detailed analysis of the underlying structure of the
system's attractor is needed to clarify the origin of this diversity, for
instance through a bifurcation analysis as in~\citet{VDDG2015}. Note that the
VDDG model is still an important tool in this hierarchy of models, since it
already contains the basic mechanisms leading to the LFV.

Another interesting finding is the change of structure of the climatologies
of the ocean gyres when choosing even or odd wave numbers ($H^{\max}$,
$M^{\max}$, $H_\text{o}^{\max}$ and $P_\text{o}^{\max}$, $P^{\max}$). Is this
feature purely associated with the convergence toward a spatially continuous
field, or does it reflect specific properties of the dynamical equations,
such as symmetries or invariance? These questions are still open and will be
the subject of a future investigation that should allow one to clarify the
best set of modes needed for the ocean description.

Finally, the aim of the model is to study the effects of specific physical
interaction mechanisms between the ocean and the atmosphere on the
mid-latitude climate, both at large and intermediate scales. The modular
design of the code of the model is adapted to such purposes, with the
possibility of implementing new components, such as oceanic active transport,
time-dependent forcings, or salinity fields.

\section{Code availability}

\textsc{Maooam} v1.0 is freely available for research purposes in the
Supplement and is also available at \url{http://github.com/Climdyn/MAOOAM}.
In addition, the code is archived at
\url{http://dx.doi.org/10.5281/zenodo.47507}. A version of the Lua
implementation which is parallelized using MPI is also available at
\url{http://github.com/Climdyn/MAOOAM/tree/mpi}. The parallelized version is
archived at \url{http://dx.doi.org/10.5281/zenodo.47510}.

\clearpage

\appendix

\section{Formulae to compute the inner products}\label{app:formulae}

In the formulae of the inner products of the atmospheric modes,
\citet{CT1987} use the following helper functions:
\begin{align}
{B_1}(u,v,w ) =& \frac{{w}+{v}}{u },\\
{B_2}(u ,v,w) =& \frac{{w}-{v}}{u },\\
&\lambda({r}) = 0~ (r~\text{even})~ \text{or}~ 1~ (r~\text{odd}), \\
S_1(u,v,w,z) =& -\frac{1}{2} (z u+w v),\\
S_2(u,v,w,z) =& \frac{1}{2} (w v-z u),\\
S_3(u,v,w,z) =& -S_1(u,v,w,z), \\
S_4(u,v,w,z) =& S_2(u,v,w,z).
\end{align}
The same notation will be used in this appendix. In what follows,
$\delta_{ij}$ is the Kronecker delta, so that $\delta_{ij}=1$ if $i=j$, and 0
otherwise. Likewise, the function $\delta(x)$ used in this appendix is
defined as
\begin{align}
\delta(x)=
    \begin{cases}
      1, & \text{if}\ x=0 \\
      0, & \text{otherwise.}
    \end{cases}
\end{align}
Using these functions, the various coefficients of the model are calculated,
starting with the internal atmosphere coefficients.

\subsection{Atmospheric coefficients}
\label{sec:atmo-coef}

In the following, we consider the ordering of the basis function used in
Eqs.~(\ref{eq:psidec})--(\ref{eq:Tdec}). For the sake of clarity, we add an
extra informative upper index specifying the type of the atmospheric function
in the definitions below. However, the inner products are completely defined
by the lower indices alone. The atmospheric functions are thus noted:
\begin{equation}
  \label{eq:afuncdef}
  F_i^\alpha(x^\prime,y^\prime) = \left\{
  \begin{array}{lc}
    \sqrt{2}\, \cos(P_i y') & \mathrm{if} ~\alpha=A, \\
    2\cos(M_i nx')\, \sin(P_i y') & \mathrm{if} ~\alpha=K, \\
    2 \sin(H_i nx')\, \sin(P_i y') & \mathrm{if}~ \alpha=L,
  \end{array}
  \right.
\end{equation}
and the oceanic functions
\begin{equation}
  \label{eq:ofuncdef}
  \phi_i(x^\prime,y^\prime) = 2\sin(\frac{H_{\text{o},i} n}{2} x')\, \sin(P_{\text{o},i} y').
\end{equation}

\subsubsection{The $a_{i,j}$ coefficients}

These coefficients correspond to the eigenvalues of the Laplacian operator
acting on the spectral expansion basis functions:
\begin{equation}
  \label{eq:aij}
  a_{i,j}^{\alpha,\beta} = \langle F_i^\alpha , \nabla^2 F_j^\beta \rangle, \quad \alpha,\beta \in
  \{A,K,L\},
\end{equation}
which are given for each case by
\begin{align}
  &a_{i,j}^{A,A} = -\delta_{ij} \, P_i^2, \label{eq:aijA} \\
  &a_{i,j}^{K,K} = -\delta_{ij} \, (n^2 M_i^2 + P_i^2), \label{eq:aijK} \\
  &a_{i,j}^{L,L} = -\delta_{ij} \, (n^2 H_i^2 + P_i^2). \label{eq:aijL}
\end{align}

\subsubsection{The $c_{i,j}$ coefficients}

These coefficients are needed to evaluate the contribution of the
$\beta$-terms, and only involve the $K$- and $L$-type base functions.
\begin{equation}
  \label{eq:cij}
  c_{i,j}^{\alpha,\beta} = \langle F_i^\alpha , \partial_{x^\prime} F_j^\beta \rangle, \quad \alpha,\beta \in  \{K,L\}.
\end{equation}
We have that
\begin{align}
  &c_{i,j}^{K,K} = c_{i,j}^{L,L} = 0,   \label{eq:cijKKLL} \\
  &c_{i,j}^{K,L} = M_i \, \delta(M_i-H_j) \, \delta(P_i-P_j) = -c_{j,i}^{L,K}.   \label{eq:cijKL}
\end{align}

\subsubsection{The $g_{i,j,k}$ coefficients}

These coefficients are given by
\begin{equation}
  \label{eq:gijk}
  g_{i,j,k}^{\alpha,\beta,\gamma} = \langle F_i^\alpha , J(F_j^\beta,F_k^\gamma) \rangle , \quad \alpha,\beta,\gamma \in
  \{A,K,L\},
\end{equation}
and the non-zero ones are given by
\begin{align}
&\hack{\hbox\bgroup\selectfont$\displaystyle}\begin{array}{l}
  g_{i,j,k}^{A,K,L} = -\frac{2\sqrt{2}}{\pi} \, M_j \left(\frac{B_1(P_i,P_j,P_k)^2}{B_1(P_i,P_j,P_k)^2-1} -\frac{B_2(P_i,P_j,P_k)^2}{B_2(P_i,P_j,P_k)^2-1}\right),  \\
  \quad \times \delta(M_j-H_k) \, \lambda(P_i+P_j+P_k), \\
  \end{array}\hack{$\egroup}\\
 &\hack{\hbox\bgroup\selectfont$\displaystyle} \begin{array}{l}
   g_{i,j,k}^{K,K,L} = S_1(P_j,P_k,M_j,H_k) \, \big\{ \delta(M_i-H_k-M_j) \, \delta(P_i-P_k+P_j)  \\
\quad  -\delta(M_i-H_k-M_j) \, \delta(P_i+P_k-P_j) \\
\quad   + \big[\delta(H_k-M_j+M_i) + \delta(H_k-M_j-M_i)\big] \, \delta(P_k+P_j-P_i) \big\}  \\
\quad  + S_2(P_j,P_k,M_j,H_k) \, \big\{ \delta(M_i-H_k-M_j) \, \delta(P_i-P_k-P_j) \\
\quad + \big[ \delta(H_k-M_j-M_i) + \delta(M_i+H_k-M_j)\big]  \\
\quad  \times \big[\delta(P_i-P_k+P_j)-\delta(P_k-P_j+P_i)\big]\big\}, \\
  \end{array}\hack{$\egroup}\\
 &\hack{\hbox\bgroup\selectfont$\displaystyle} \begin{array}{l}
   g_{i,j,k}^{L,L,L}= S_3(P_j,P_k,H_j,H_k) \, \big\{\delta(H_k+H_j-H_i) \, \delta(P_k-P_j+P_i)  \\
\quad  + \big[\delta(H_k-H_j-H_i)-\delta(H_k-H_j+H_i)\big] \, \delta(P_k+P_j-P_i)  \\
\quad   - \delta(H_k+H_j-H_i) \, \delta(P_k-P_j-P_i) \big\} \\
\quad  + S_4(P_j,P_k,H_j,H_k) \, \{\delta(H_k+H_j-H_i) \, \delta (P_k-P_j-P_i)  \\
 \quad  + \big[ \delta(H_k-H_j+H_i) - \delta(H_k-H_j-H_i) \big] \\
\quad \times \big[\delta(P_k-P_j-P_i)-\delta(P_k-P_j+P_i)\big]\big\}~
\mathrm{for} ~ k \geq j \geq
  i,\\
  \end{array}\hack{$\egroup}
\end{align}
where we have used the functions defined at the beginning of this appendix.
All the other permutations can be obtained thanks to
\begin{equation}
  \label{eq:gperm}
  g_{i,j,k}^{\alpha,\beta,\gamma} = - g_{j,i,k}^{\beta,\alpha,\gamma} = g_{k,i,j}^{\gamma,\alpha,\beta} = g_{j,k,i}^{\beta,\gamma,\alpha}.
\end{equation}

\subsubsection{The $b_{i,j,k}$ coefficients}

These coefficients are given by
\begin{equation}
  \label{eq:bijk}
  b_{i,j,k}^{\alpha,\beta,\gamma} = \langle F_i^\alpha , J(F_j^\beta, \nabla^2 F_k^\gamma) \rangle  , \quad \alpha,\beta,\gamma \in  \{K,L\}.
\end{equation}
Therefore we obtain
\begin{equation}
  \label{eq:bijkdef}
  b_{i,j,k}^{\alpha,\beta,\gamma} = a_{k,k}^{\gamma,\gamma} \langle F_i^\alpha , J(F_j^\beta,  F_k^\gamma) \rangle = a_{k,k}^{\gamma,\gamma} \, g_{i,j,k}^{\alpha,\beta,\gamma}.
\end{equation}

\subsubsection{The $s_{i,j}$ coefficients}

These coefficients encode the inner products between the atmospheric and
oceanic basis functions:
\begin{equation}
  \label{eq:sijdef}
  s_{i,j}^\alpha = \langle F_i^\alpha , \phi_j \rangle , \quad \alpha \in  \{A,K,L\},
\end{equation}
which gives
\begin{align}
 & s_{i,j}^A  =  8 \, \sqrt{2} \, P_{\text{o},j} \, \frac{\lambda(H_{\text{o},j}) \, \lambda(P_{\text{o},j}+P_i)}{\pi^2 \, H_{\text{o},j} \, (P_{\text{o},j}^2 - P_i^2) }, \\
 & s_{i,j}^K  = 4 \, H_{\text{o},j} \, \frac{\lambda(2M_i+H_{\text{o},j}) \, \delta(P_{\text{o},j}-P_i)}{\pi  (H_{\text{o},j}^2 - 4 M_i^2) }, \\
 & s_{i,j}^L  =  \delta(P_{\text{o},j}-P_i) \, \delta(2 H_i - H_{\text{o},j}).
\end{align}

\subsubsection{The $d_{i,j}$ coefficients}

These coefficients are related to the forcing of the ocean on the atmosphere.
They are given by the formula
\begin{equation}
  \label{eq:dijdef}
  d_{i,j}^\alpha = \langle F_i^\alpha , \nabla^2 \phi_j \rangle = M_{j,j} \, s_{i,j}^\alpha , \quad \alpha \in
  \{A,K,L\},
\end{equation}
where the $M_{j,j}$ are given by the eigenvalues of the Laplacian operator
acting on the oceanic basis functions (see next section).

\subsection{Oceanic coefficients}
\label{sec:oc-coef}

\subsubsection{The $K_{i,j}$ coefficients}

These coefficients are related to the forcing of the atmosphere on the ocean.
They are given by
\begin{equation}
  \label{eq:Kijdef}
  K_{i,j}^{\alpha} = \langle  \phi_i , \nabla^2 F_j^\alpha \rangle = a_{j,j}^{\alpha,\alpha} \, s_{j,i}^\alpha , \quad \alpha \in  \{A,K,L\}.
\end{equation}

\subsubsection{The $M_{i,j}$ coefficients}

These coefficients identify with the eigenvalues of the Laplacian acting on
the oceanic basis functions:
\begin{equation}
  \label{eq:Mijdef}
  M_{i,j} = \langle \phi_i , \nabla^2 \phi_j \rangle = - \delta_{ij} (n^2 H_{\text{o},i}^2 / 4 + P_{\text{o},i}^2).
\end{equation}

\subsubsection{The $N_{i,j}$ coefficients}

These coefficients are needed to evaluate the contribution of the
$\beta$-terms and are given by
\begin{align}
  \label{eq:Nijdef}
  N_{i,j} &= \langle \phi_i , \partial_{x^\prime} \phi_j \rangle \\
  &= -2 \, n \, H_{\text{o},i} \, H_{\text{o},j} \frac{\delta(P_{\text{o},i}-P_{\text{o},j}) \,
  \lambda(H_{\text{o},i} + H_{\text{o},j})}{\pi \,
  (H_{\text{o},j}^2-H_{\text{o},i}^2)}.\nonumber
\end{align}

\subsubsection{The $O_{i,j,k}$ coefficients}

These coefficients are given by
\begin{equation}
  \label{eq:Oijk}
  O_{i,j,k} = \langle \phi_i , J(\phi_j,\phi_k) \rangle
\end{equation}
with
\begin{align}
&\hack{\hbox\bgroup\selectfont$\displaystyle}\begin{array}{l}
  O_{i,j,k}  = \frac{n}{2} \, \Big[ \, S_3(P_{\text{o},j},P_{\text{o},k},H_{\text{o},j},H_{\text{o},k})  \big\{[\delta(H_{\text{o},k}-H_{\text{o},j}-H_{\text{o},i}) \\
\quad  - \delta(H_{\text{o},k}-H_{\text{o},j}+H_{\text{o},i})] \, \delta(P_{\text{o},k}+P_{\text{o},j}-P_{\text{o},i}) \\
\quad + \delta(H_{\text{o},k}+H_{\text{o},j}-H_{\text{o},i})\, [\delta(P_{\text{o},k}-P_{\text{o},j}+P_{\text{o},i})   \\
\quad - \delta(P_{\text{o},k}-P_{\text{o},j}-P_{\text{o},i})]\big\} + \, S_4(P_{\text{o},j},P_{\text{o},k},H_{\text{o},j},H_{\text{o},k}) \,  \\
\quad \times \big\{[\delta(H_{\text{o},k}+H_{\text{o},j}-H_{\text{o},i})\, \delta(P_{\text{o},k}-P_{\text{o},j}-P_{\text{o},i})]\\
\quad + [\delta(H_{\text{o},k}-H_{\text{o},j}+H_{\text{o},i}) - \delta(H_{\text{o},k}-H_{\text{o},j}-H_{\text{o},i})] \\
\quad \times [\delta(P_{\text{o},k}-P_{\text{o},j}-P_{\text{o},i}) -
\delta(P_{\text{o},k}-P_{\text{o},j}+P_{\text{o},i})]\big\} \, \Big]\\
\quad \mathrm{for} ~k \geq j \geq i.\\
\end{array}\hack{$\egroup}
\end{align}

\subsubsection{The $C_{i,j,k}$ coefficients}

These coefficients are given by
\begin{equation}
  \label{eq:Cijk}
  C_{i,j,k} = \langle \phi_i , J(\phi_j,\nabla^2 \phi_k) \rangle = M_{k,k} \, O_{i,j,k}.
\end{equation}

\subsubsection{The $W_{i,j}$ coefficients}

These coefficients are related to the short-wave radiative forcing of the
ocean and are given by
\begin{equation}
  \label{eq:Wij}
  W_{i,j}^{\alpha} = \langle \phi_i, F_j^\alpha  \rangle = s_{j,i}^\alpha, \quad \alpha \in  \{A,K,L\}.
\end{equation}

\section{Definition of the tensor $\mathcal{T}_{i,j,k}$}
\label{app:tensor}

The system of non-dimensionalized ODEs for the model variables is encoded in
the model tensor $\mathcal{T}_{i,j,k}$, of which the complete definition is
given in this appendix. Tensor elements that are not listed below are equal
to zero. To alleviate the notations, we use a shorthand notation for the
indices of the different variables,
\begin{align}
& \psi_i   = i  &(1 \leq i \leq n_\text{a}),~\\
& \theta_i = i+n_\text{a} &(1 \leq i \leq n_\text{a}),~ \nonumber\\
& \Psi_i       = i+2n_\text{a} &(1 \leq i \leq n_\text{o}),~ \nonumber\\
& \Theta_i = i+2n_\text{a}+n_\text{o} &(1 \leq i \leq n_\text{o}).\nonumber
\end{align}
Furthermore, we suppress the upper indices which indicate the atmospheric
function types but are otherwise not needed to unambiguously specify the
inner products.

\subsection{Atmosphere equations}

The components of the tensor for the atmosphere streamfunction are given by
\begin{align}
&\mathcal{T}_{\psi_i,\psi_j,0}=  - \frac{c_{i,j}\; \beta'}{a_{i,i}}-\frac{k_d}{2}\delta_{i,j}, \\
&\mathcal{T}_{\psi_i,\theta_j,0} = \frac{k_d}{2}\delta_{i,j},\nonumber \\
&\mathcal{T}_{\psi_i,\psi_j,\psi_k} = \mathcal{T}_{\psi_i,\theta_j,\theta_k} = -\frac{b_{i,j,k}}{a_{i,i}},\nonumber\\
&\mathcal{T}_{\psi_i,\Psi_j,0} = \frac{k_d\; d_{i,j}}{2a_{i,i}}.\nonumber
\end{align}

The atmospheric temperature equations are determined by the tensor elements
\begin{align}
&\mathcal{T}_{\theta_1,0,0} = \frac{C'_\text{a}}{1 - a_{1,1}\;\sigma_0},\\
&\mathcal{T}_{\theta_i,\psi_j,0} = \frac{a_{i,j}\; k_d \;\sigma_0}{2 a_{i,i}\;\sigma_0 - 2},\nonumber\\
&\hack{\hbox\bgroup\selectfont$\displaystyle}\mathcal{T}_{\theta_i,\theta_j,0}
= -  \frac{\sigma_0 \left( 2\;c_{i,j} \;\beta' + a_{i,j}(k_d+4\; k_d')\right) - 2\;(S_\text{B,a}' + \lambda_\text{a}')\delta_{i,j}}{2\;a_{i,i}\;\sigma_0 - 2},\nonumber \hack{$\egroup}\\
&\mathcal{T}_{\theta_i,\psi_j,\theta_k} = \frac{g_{i,j,k}-b_{i,j,k}\;\sigma_0 }{a_{i,i}\;\sigma_0 - 1},\nonumber\\
&\mathcal{T}_{\theta_i,\theta_j,\psi_k} = \frac{b_{i,j,k}\;\sigma_0}{1-a_{i,i}\sigma_0}, \nonumber\\
&\mathcal{T}_{\theta_i,\Psi_j,0} = \frac{ k_d\; d_{i,j} \;\sigma_0}{2-2 a_{i,i}\;\sigma_0}, \nonumber\\
&\mathcal{T}_{\theta_i,\Theta_j,0} = s_{i,j} \frac{2 \;S_\text{B,o}' +
\lambda_\text{a}'}{2 - 2\;a_{i,j}\;\sigma_0},\nonumber
\end{align}
where we used the non-dimenionalized quantities
\begin{align*}
&\beta' = \beta L/f_0, \\
&C'_\text{a} = C_\text{a} R/(2 \gamma_\text{a} f_0^3 L^2), \\
&\sigma_0 = \sigma \Delta p^2/(2 L^2 f_0^2), \\
&\lambda'_\text{a} = \lambda/(\gamma_\text{a} f_0), \\
&S_\text{B,o}' =  2 \epsilon_\text{a} \sigma_\text{B} \left(T_\text{o}^0\right)^3/( \gamma_\text{a} f_0), \\
&S_\text{B,a}' = 8 \epsilon_\text{a} \sigma_\text{B}
\left(T_\text{a}^0\right)^3 /(\gamma_\text{a} f_0).
\end{align*}

\subsection{Ocean equations}

The components of the tensor for the ocean streamfunction are
\begin{align}
&\mathcal{T}_{\Psi_i,\psi_j,0} = -\mathcal{T}_{\Psi_i,\theta_j,0} = \frac{K_{i,j} \;d'}{M_{i,i} + \gamma},  \\
&\mathcal{T}_{\Psi_i,\Psi_j,0} = -\frac{N_{i,j}\; \beta' + \delta_{i,j}\; M_{i,i}\;(r' + d')}{M_{i,i} + \gamma},\nonumber \\
&\mathcal{T}_{\Psi_i,\Psi_j,\Psi_k} = - \frac{C_{i,j,k}}{M_{i,i} +
\gamma},\nonumber
\end{align}
with $\gamma = -L/L_\text{R}$, $d' = d/ f_0$, and $r' = r/ f_0$.

Finally, the equations for the ocean temperature are determined by
\begin{align}
&\mathcal{T}_{\Theta_i,0,0} = C_\text{o}'\; W_{i,1},\\
&\mathcal{T}_{\Theta_i,\theta_j,0} = W_{i,j}(2\;\lambda_\text{o}' + \sigma_\text{B,a}'),\nonumber\\
&\mathcal{T}_{\Theta_i,\Theta_j,0} = -\delta_{i,j} (\lambda_\text{o}' + \sigma_\text{B,o}'),\nonumber\\
&\mathcal{T}_{\Theta_i,\Psi_j,\Theta_k} = -O_{i,j,k},\nonumber
\end{align}
where the following non-dimensionalized quantities are used:
\begin{align*}
&C'_\text{o} = C_\text{o} R/(\gamma_\text{a} f_0^3 L^2), \\
&\lambda'_\text{o} = \lambda/(\gamma_\text{o} f_0), \\
&\sigma_\text{B,o}' = 4 \sigma_\text{B} \left(T_\text{o}^0\right)^3/( \gamma_\text{o} f_0), \\
&\sigma_\text{B,a}'= 8 \epsilon_\text{a} \sigma_\text{B}
\left(T_\text{a}^0\right)^3 /(\gamma_\text{o} f_0).
\end{align*}

\begin{acknowledgements}
This work is partly supported by the Belgian Federal Science Policy Office
under contract BR/121/A2/STOCHCLIM. The figures and videos have been prepared
with the Matplotlib software~\citep{H2007}.\hack{\newline} \hack{\newline}
Edited by: O.~Marti\hack{\newline} Reviewed by: T. Sengul and one anonymous
referee
\end{acknowledgements}

\begin{figure}[t]
\includegraphics[width=\textwidth]{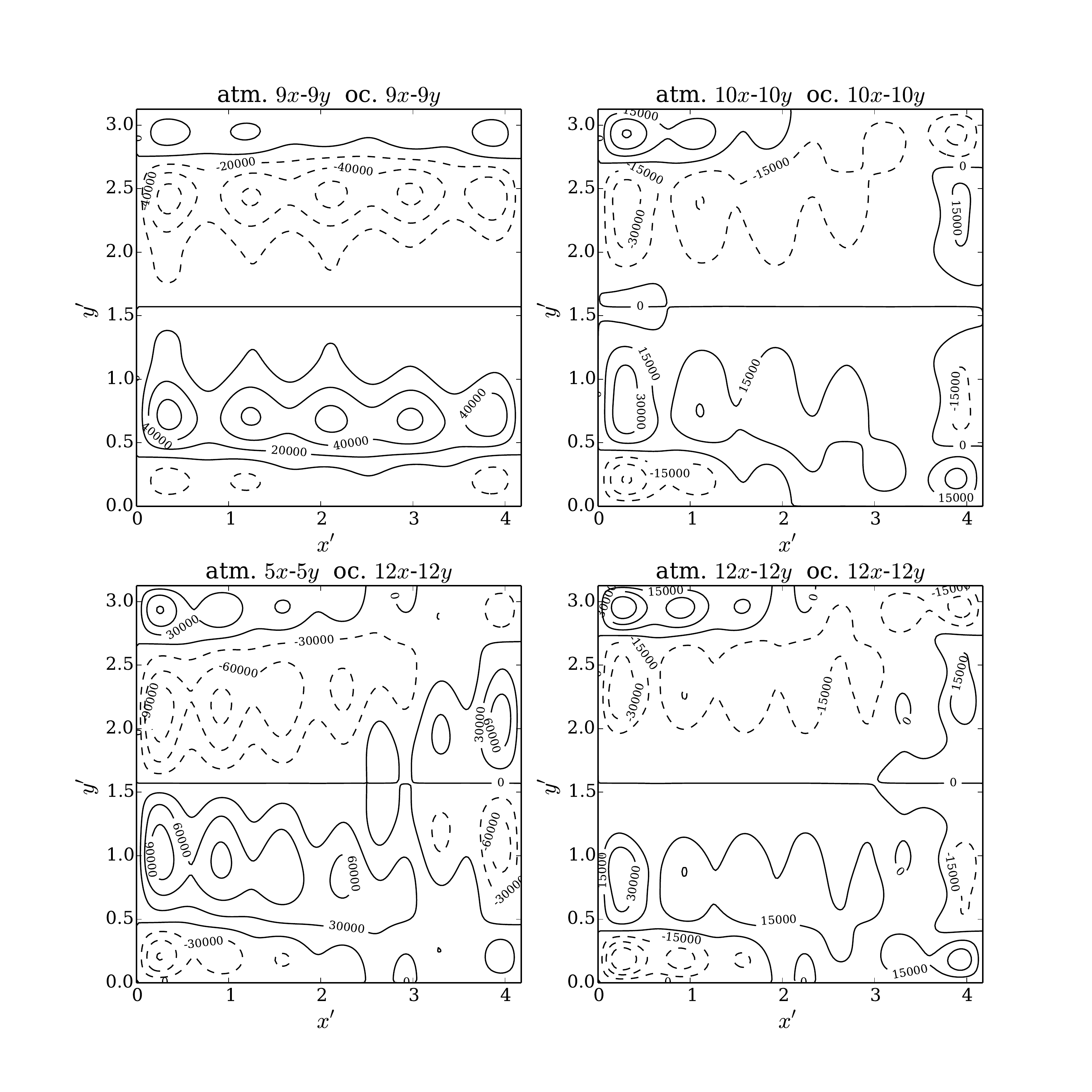}
\caption{Climatologies for the oceanic streamfunction $\psi_\text{o}$ field
(m$^2$\,s$^{-1}$) presented on the non-dimensional model domain for the
high-resolution runs displayed in Fig.~\ref{fig:var_A_GRL2_high}.
\label{fig:clim_A_GRL2_high}}
\end{figure}

\begin{table}[p]
 \caption{Values of the parameters of the model that are used in the analyses of Sect.~\ref{sec:dynamics}.}
   \begin{tabular}{lllr}
    \tophline
    Parameter (unit) & Value & Parameter (unit) & Value \\
    \hhline
    $n = 2 L_y / L_x$         & $1.5$           & $L_\text{R}$ (km)                  & $19.93$\\
    $L_y = \pi L$  (km)       & $5.0 \times 10^3$     & $\rho$ (kg m$^{-3}$)        & $1000$ \\
    $f_0$ (s$^{-1}$)          & $1.032 \times 10^{-3}$ & $\sigma_\text{B}$ (W\,m$^2$\,K$^{-4}$) & $5.6 \times 5.6 10^{-8}$\\
    $\lambda$ (W\,m$^{-2}$\,K$^{-1}$) & $15.06$         & $\sigma$ (m$^2$\,s$^{-2}$\,Pa$^{-2}$) & $2.16 \times 10^{-6}$\\
    $r$ (s$^{-1}$)            & $1.0 \times 10^{-7}$        & $\beta$ (m$^{-1}$\,s$^{-1}$)  & $1.62 \times 10^{-11}$ \\
    $d$ (s$^{-1}$)            & $1.1 \times 10^{-7}$  & $R$ (J\,kg$^{-1}$\,K$^{-1}$)  & $287$ \\
    $C_\text{o}$ (W\,m$^{-2}$)     & $310$                 & $\gamma_\text{o}$ (J\,m$^{-2}$\,K$^{-1}$) & $5.46 \times 10^8$ \\
    $C_\text{a}$ (W\,m$^{-2}$)     & $C_\text{o}/3$            & $\gamma_\text{a}$ (J\,m$^{-2}$\,K$^{-1}$) & $1.0 \times 10^7$ \\
    $k_d$  (s$^{-1}$)         & $3.0 \times 10^{-6}$ & $T_\text{a}^0$ (K)                  & $289.30$ \\
    $k_d^\prime $  (s$^{-1}$)       & $3.0 \times 10^{-6}$ & $T_\text{o}^0$ (K)                  & $301.46$ \\
    $h$ (m)                   & $136.5$         &  $\epsilon_\text{a}$         & $0.7$ \\
    \bottomhline
  \end{tabular}
  \label{tab:params}
\end{table}

\begin{table}[p]
\caption{Number of variables, transient time, and effective runtime of the
runs (in years).} 
\begin{tabular}{lrrr}
    \tophline
    Resolution & No. of  & Transient  & Eff. runtime \\
 &  variables & ($y$) &  ($y$) \\
    \hhline
    atm. $2x$--$2y$ oc. $2x$--$4x$ & $36$ & 30\,726.5 & 92\,179.6 \\
    atm. $2x$--$4y$ oc. $2x$--$4x$ & $56$ & 30\,726.5 & 92\,179.6 \\
    atm. $3x$--$3y$ oc. $3x$--$3x$ & $60$ & 30\,726.5 & 92\,179.6 \\
    atm. $4x$--$4y$ oc. $4x$--$4x$ & $104$ & 30\,726.5 & 92\,179.6 \\
    \hhline
    atm. $5x$--$5y$ oc. $5x$--$5x$ & $160$ & 30\,726.5 & 92\,179.6 \\
    atm. $6x$--$6y$ oc. $6x$--$6x$ & $228$ & 30\,726.5 & 92\,179.6 \\
    atm. $7x$--$7y$ oc. $7x$--$7x$ & $308$ & 30\,726.5 & 92\,179.6 \\
    atm. $8x$--$8y$ oc. $8x$--$8x$ & $400$ & 30\,726.5 & 92\,179.6 \\
    \hhline
    atm. $9x$--$9y$ oc. $9x$--$9y$ & $504$ & 30\,726.5 & 92\,179.6 \\
    atm. $10x$--$10$ oc. $10x$--$10y$ & $620$ & 30\,726.5 & 92\,179.6 \\
    atm. $5x$--$5y$ oc. $12x$--$12y$ & $398$ & 15\,363.3 & 92\,179.6 \\
    atm. $12x$--$12y$ oc. $12x$--$12y$ & $888$ & 15\,363.3 & 74\,972.7 \\
    \bottomhline
  \end{tabular}
     \label{tab:runtime}
\end{table}

\end{document}